\documentclass[sigconf]{acmart}
\AtBeginDocument{%
  }

\copyrightyear{2026}
\acmYear{2026}
\setcopyright{cc}
\setcctype{by}
\acmConference[CCS '26] {Proceedings of the 2026 ACM SIGSAC Conference on Computer and Communications Security}{November 15--19, 2026}{The Hague, Netherlands.}
\acmBooktitle{Proceedings of the 2026 ACM SIGSAC Conference on Computer and Communications Security (CCS '26), November 15--19, 2026, The Hague, Netherlands}
\acmISBN{979-8-4007-2871-6/2026/11}
\acmDOI{10.1145/3830454.3846584}

\usepackage{enumitem}
\usepackage{subcaption} 
\usepackage{multirow}
\usepackage[most]{tcolorbox}
\usepackage{xcolor}
\usepackage{amsthm}
\usepackage{algorithm}
\usepackage{algpseudocode}
\usepackage{array}
\usepackage{url}
\usepackage{booktabs}
\usepackage{hyperref}
\usepackage{tabularx}
\usepackage{listings}

\usepackage[flushleft]{threeparttable}

\newcommand{\tool}{\textsc{PVParser}}

\begin{document}

\title{Recovering Process Variables from Industrial Network Traffic via Search-Based Optimization}

\author{Chuan Sheng}
\affiliation{%
  \institution{Shenyang Institute of Automation, Chinese Academy of Sciences}
  \city{Shenyang}
  \state{Liaoning}
  \country{China}
  }
\email{shengchuan@sia.cn}

\author{Shan Jiang}
\authornote{Corresponding author.}
\affiliation{%
  \institution{Swinburne University of Technology}
  \city{Melbourne}
  \state{Victoria}
  \country{Australia}
  }
\email{shanjiang@swin.edu.au}



\author{Jianming Zhao}
\affiliation{%
  \institution{Shenyang Institute of Automation, Chinese Academy of Sciences}
  \city{Shenyang}
  \state{Liaoning}
  \country{China}
  }
\email{zhaojianming@sia.cn}

\author{Yu Yao}
\affiliation{%
  \institution{Northeastern University}
  \city{Shenyang}
  \state{Liaoning}
  \country{China}
  }
\email{yaoyu@mail.neu.edu.cn}





\begin{abstract}
Process variables (PVs) provide the process evidence needed for process-aware security monitoring in industrial cyber-physical systems (CPSs). However, existing supervisory infrastructures expose only the subset of PV values recorded by historians, leaving many additional runtime PV values unobserved. 
To address this incomplete process visibility, we study the problem of recovering PV fields and their semantics directly from raw industrial network traffic through protocol reverse engineering (PRE).
In this setting, existing PRE methods face two practical challenges: PV-carrying communication is mixed with heterogeneous runtime traffic, and PV-carrying payloads are often long and deployment-specific. Mixed runtime traffic obscures the PV-carrying communication paths, while long payloads create a vast segmentation space in which early segmentation errors can propagate and corrupt the recovery of later fields under sequential inference. 
In this paper, we formulate the recovery of PV fields from raw network traffic as a search-based optimization problem. Our key insight is that non-sequentially identifying correct segmentations in such a vast segmentation space can be cast as an optimization problem and addressed by searching for near-optimal solutions. We propose \tool{} to approach this goal. \tool{} first reduces the search space by identifying the PV-carrying payloads from network traffic via a periodic pattern detection mechanism. It then employs a modified Monte Carlo Tree Search to explore near-optimal segmentations, reducing error propagation from incorrect early boundary decisions. Experiments on three representative industrial CPS datasets demonstrate that \tool{} achieves high accuracy and F1-score in PV-carrying payload localization and PV field inference, outperforming six state-of-the-art PRE approaches by a significant margin. We further demonstrate that recovering missing process visibility can strengthen downstream attack detection in process-aware security monitoring.

\noindent\textbf{Full Version.}
The full version of this paper, including appendices, is available at
\href{https://arxiv.org/abs/2608.16403}{\textbf{https://arxiv.org/abs/2608.16403}}.
\end{abstract}




\maketitle

\section{Introduction}
Attacks against physical processes in industrial cyber-physical systems (CPSs) have already led to severe real-world consequences, including incidents such as Stuxnet~\cite{kushner2013real} and BlackEnergy~\cite{case2016analysis}. 
Defending against such attacks increasingly relies on process-aware security monitoring, which uses process variables (PVs), such as sensor measurements and actuator states, to reason about whether the physical process follows expected runtime behavior~\cite{xu2024physcout, cai2024ics, sun2024secure}. 
In operational deployments, defenders mainly obtain PV evidence from supervisory components, such as SCADA servers and historians~\cite{ur2024process}. 
However, these components provide only partial and sampled PV records, while many runtime PV values remain in lower-level industrial communication and are unavailable to defenders. 
This creates an asymmetric visibility gap: defenders observe only partial PV evidence from supervisory records, while attacks may affect a broader set of runtime PV values in lower-level industrial communication. 
This gap weakens process-aware security monitoring against stealthy process manipulation and split-view deception attacks~\cite{urbina2016limiting, pu2024cormand2}. 
Recovering PVs from industrial network traffic therefore helps bridge this gap by restoring missing runtime PV evidence for process-aware security monitoring.

This paper addresses the problem of recovering PV fields and their semantics directly from raw industrial network traffic.
This problem is particularly relevant in legacy or vendor-maintained industrial CPS deployments, where defenders have access to mirrored traffic and historian records, but incomplete, outdated, or unavailable decoding artifacts leave important runtime PV evidence encoded in deployment-specific payloads and inaccessible to security monitoring.
In these settings, protocol reverse engineering (PRE) provides a natural approach for recovering deployment-specific PV fields from observed traffic.
However, PV recovery from raw industrial traffic faces two technical challenges.
First, PV-carrying communication is mixed with heterogeneous industrial traffic, where monitoring, control, and diagnostic messages are interleaved and only a subset carries PV values.
Second, PV fields are encoded in long deployment-specific payloads, creating a vast candidate segmentation space for inferring field boundaries and semantics.

Under these conditions, existing PRE approaches fall short of directly recovering PVs from raw industrial traffic.
Active approaches deliberately perturb the control loop to trigger PV changes and correlate them with observed traffic~\cite{verma2021can, cai2022seminer, yu2022towards, lin2024bycan}. 
Such interventions are infeasible in operational CPSs, where modifying controller behavior can compromise safety~\cite{stouffer2011guide}. 
Passive methods avoid intervention, but generally assume that target message traces are available and rely on heuristic-driven sequential inference to recover field boundaries and semantics~\cite{qin2024reverse, huang2024finebid}.
These assumptions limit passive methods for PV recovery from raw industrial traffic: mixed runtime traffic prevents direct access to PV-carrying payloads, while the vast segmentation space of long deployment-specific payloads makes sequential inference prone to incorrect early boundary decisions.
Such errors can propagate through the remaining segmentation process and prevent reliable recovery of later PV fields~\cite{ye2021netplier, chandler2023binaryinferno}.

To overcome these limitations, we formulate PV recovery from raw industrial traffic as a search problem over communication paths and payload segmentations. 
Specifically, we first narrow the search over traffic to PV-carrying communication paths and then formulate PV field inference as a search-based optimization problem over candidate payload segmentations. 
The key insight is that PV field inference should optimize the segmentation of the whole payload rather than decide each field boundary independently. 
By searching for near-optimal payload-level segmentations, inference can avoid prematurely choosing incorrect early field boundaries.

Based on this formulation, we propose \tool{}, a two-stage framework for PV recovery from raw industrial traffic. 
The first stage localizes PV-carrying payloads from heterogeneous runtime traffic, and the second stage infers PV fields within the localized payloads through search-based optimization. 
To localize PV-carrying payloads, \tool{} leverages the observation that PV-carrying communication often appears as recurrent request-response interactions, especially periodic polling. 
By discovering such periodic patterns, \tool{} recovers PV-carrying communication paths from heterogeneous runtime traffic and extracts the corresponding payloads for field inference. 
To infer fields within the localized payloads, \tool{} employs a modified Monte Carlo Tree Search (MCTS) algorithm~\cite{browne2012survey} that jointly evaluates local boundary plausibility and global structural consistency. 
This allows \tool{} to compare plausible candidate segmentations at the payload level and limit error propagation from incorrect early field boundaries, enabling accurate PV field recovery.

We implement \tool{} and evaluate it on three representative industrial CPS datasets: SCADA-N~\cite{lemay2016providing}, SWaT~\cite{goh2016dataset}, and WADI~\cite{ahmed2017wadi}. 
For PV field inference, we compare \tool{} with six state-of-the-art PRE methods: NetPlier~\cite{ye2021netplier}, BinaryInferno~\cite{chandler2023binaryinferno}, CAN-D~\cite{verma2021can}, ByCAN~\cite{lin2024bycan}, DP-Reverser~\cite{yu2022towards}, and SeMiner~\cite{cai2022seminer}. 
\tool{} improves accuracy by over 0.38 and F1-score by over 0.24 against the strongest baseline on SWaT and WADI.
We further demonstrate the security utility of recovered process visibility in downstream process-aware attack detection. 
Across four representative attack scenarios, process-aware detectors augmented with recovered PV evidence improve F1-score and accuracy by up to 0.84 and 0.46 over the original detectors, respectively.

We summarize our contributions as follows:

\begin{itemize}[leftmargin=*]

    \item To the best of our knowledge, we present the first framework for recovering PVs directly from raw industrial network traffic, restoring missing runtime PV evidence to strengthen process-aware security monitoring.
    
    \item We introduce a PV-carrying payload localization mechanism that exploits recurrent communication patterns, especially periodic polling, to recover PV-carrying communication paths from heterogeneous runtime traffic.

    \item We introduce a search-based optimization algorithm for PV field inference based on a modified MCTS, which jointly evaluates local boundary plausibility and global structural consistency to reliably recover fields from long deployment-specific payloads.

    \item We implement \tool{} and evaluate it on three representative industrial CPS datasets, demonstrating superior PV recovery performance over six state-of-the-art PRE methods and improved process-aware attack detection. The tool is released~\cite{pvparser_artifact}.
\end{itemize}

\begin{figure}[!t]
	\centering
        \includegraphics[width=0.95\columnwidth]{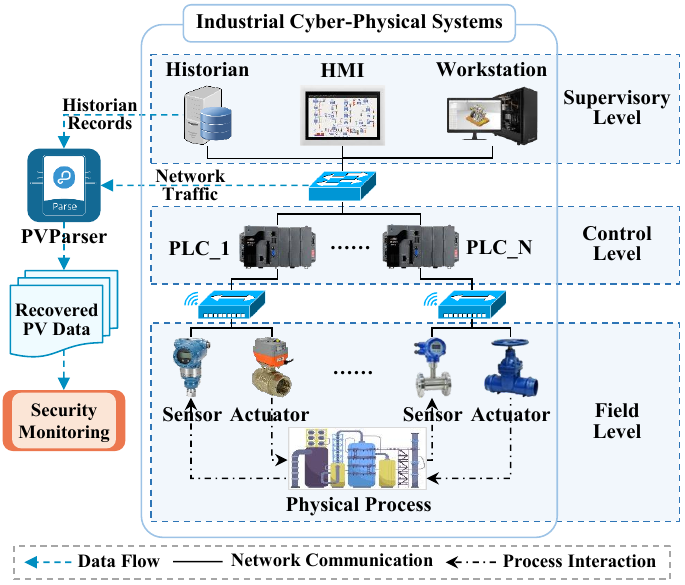}
    \caption{Process visibility from supervisory records and network traffic in industrial CPSs.}
	\label{fig:fig2.1}
\end{figure}

\section{Background and Problem Scope}
This section provides the background and problem scope for PV recovery in industrial CPSs. It frames PV recovery as a way to extend process visibility beyond supervisory-level records and to strengthen process-aware security monitoring.

\subsection{Process Visibility in Industrial CPSs}
In this paper, process visibility refers to the view of system operation reflected by PVs. This section explains how such visibility is obtained from supervisory-level PV records and why it matters for process-aware security monitoring.

\subsubsection{Supervisory-Level Process Observation}
As illustrated in Figure~\ref{fig:fig2.1}, industrial CPSs are typically organized into the supervisory, control, and field levels~\cite{stouffer2011guide}. The supervisory level hosts components such as human-machine interfaces (HMIs) and historians for process monitoring and data recording, whereas the control level hosts PLCs and other controllers that maintain and transmit runtime PVs. In such systems, system operation is primarily observed at the supervisory level through PVs~\cite{ur2024process}.

\begin{figure}[!t]
	\centering
        \includegraphics[width=0.9\columnwidth]{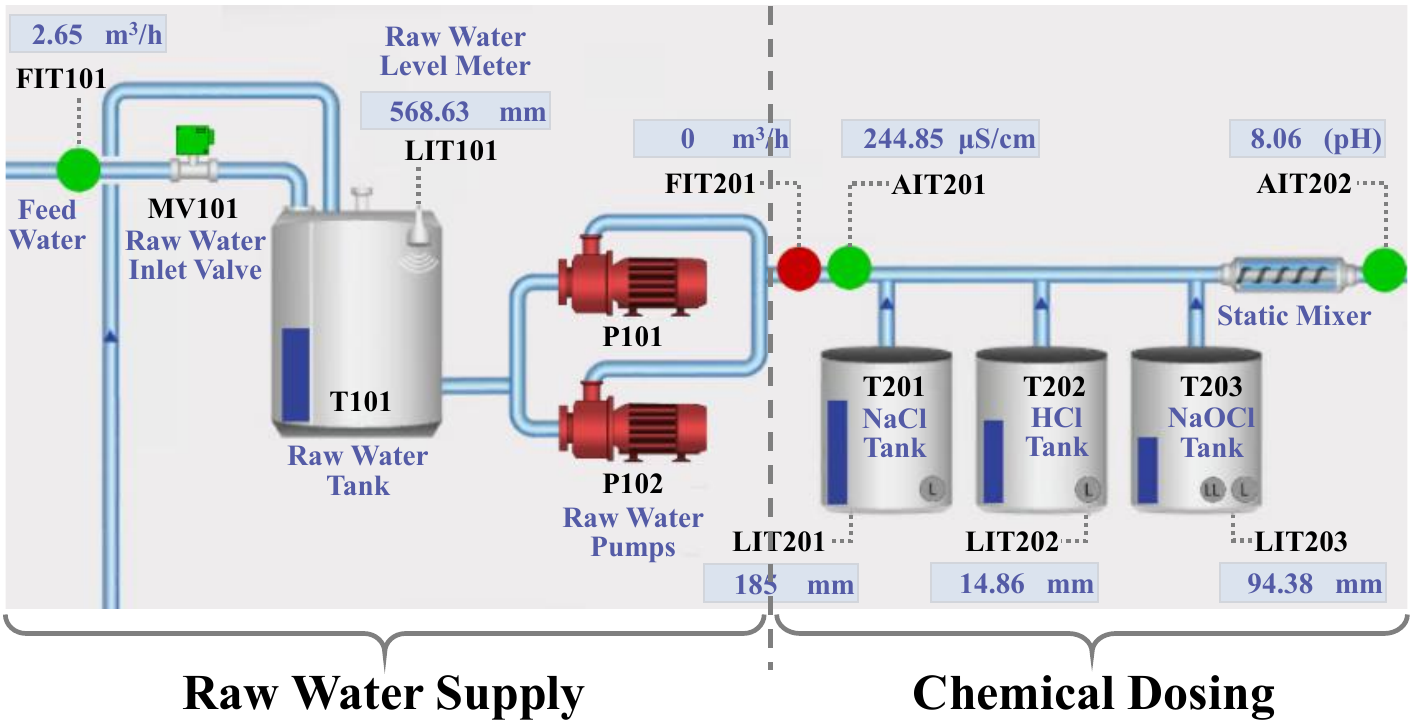}
    \caption{Representative process variables from two subprocesses of a water treatment system~\cite{goh2016dataset}.}
	\label{fig:fig2.2}
\end{figure}

PVs are sensor measurements and actuator states that characterize the runtime state of a physical process. Figure~\ref{fig:fig2.2} shows representative PVs from SWaT, including sensor measurements such as LIT101 and FIT101 and actuator states such as MV101 and P101. In this paper, supervisory-level PVs are the PV values available in historian or HMI records. They are supervisory-level observations of the same underlying PVs rather than a separate class of PVs.

However, supervisory-level records preserve only sampled PV observations rather than all runtime PV values, and they contain only the PV data forwarded upward for monitoring and archival purposes. Many runtime PV transmissions remain at the control level and therefore do not appear in these records. This gap is substantial: in the SWaT testbed~\cite{goh2016dataset}, for example, over 95\% of PV-carrying messages occur in control-level communication rather than in supervisory-level observation. Consequently, supervisory-level PV records provide only a partial and sampled view of actual system operation.

\subsubsection{Process Variables in Process-Aware Security Monitoring}
In industrial CPSs, process-aware security monitoring detects attacks against physical processes by examining whether these processes follow expected runtime behavior~\cite{ur2024process}. PVs provide the basic evidence for this examination because they describe the runtime state of physical processes. In practice, existing process-aware detectors commonly rely on historian-observable supervisory-level PVs to characterize normal runtime behavior and identify abnormal system states~\cite{aoudi2018truth, abbas2024sain, fung2024attributions, wolsing2025gecos}.

However, relying only on supervisory-level PV records can leave process-aware detectors vulnerable in two cases. First, stealthy attacks that gradually induce process deviations may be difficult to expose from the partial and sampled supervisory-level view~\cite{urbina2016limiting}. Second, under split-view deception, an attacker may replay benign-looking PV values at the supervisory level while manipulated values are used in control-level communication~\cite{pu2024cormand2}. Both cases stem from the same limitation: detectors that rely only on supervisory-level records lack access to control-level PV values.

This limitation motivates the design of PVParser as a visibility-recovery front-end that recovers missing control-level PVs from network traffic to strengthen existing process-aware detectors. The recovered control-level view can help expose attack effects that are delayed, weakened, or absent in supervisory-level records. It can also reveal split-view deception when supervisory-level records remain benign-looking while control-level communication carries manipulated PV values. Section~\ref{sec:security_eval} later evaluates how recovered control-level PVs support downstream attack detection.

\begin{figure}[!t]
	\centering
        \includegraphics[width=0.9\columnwidth]{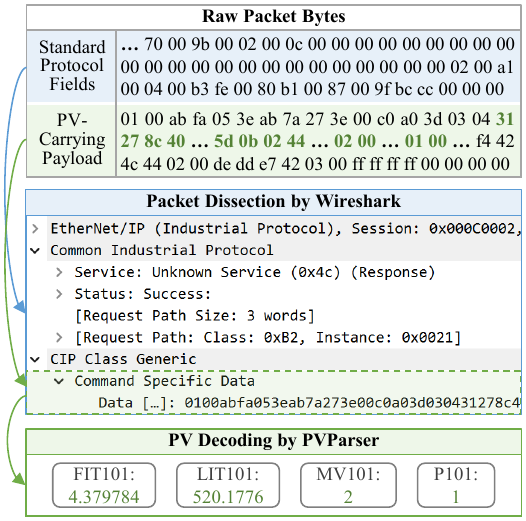}
    \caption{Comparison of standard packet dissection and deployment-specific PV decoding for the same SWaT EtherNet/IP/CIP packet.}
	\label{fig:wireshark}
\end{figure}

\subsection{PV-Carrying Communication and Payloads}
\label{sec:pv_comm}

In industrial CPSs, PV-carrying communication is mixed with heterogeneous runtime traffic between supervisory and control components. This traffic supports process monitoring, control commands, and fault diagnostics, but only part of it carries PV values. As a result, PV-carrying messages are interleaved with other industrial messages. In many deployments, they appear as recurrent request-response interactions, especially periodic polling between supervisory components and controllers~\cite{lin2019timing}. These interactions form PV transmission paths, with PV values embedded in the corresponding payloads. Therefore, recovering PVs from mixed industrial traffic first requires identifying PV transmission paths and localizing the PV-carrying payloads on these paths.

Once PV-carrying payloads are localized, standard protocol dissectors can parse the standard protocol fields but may not resolve the deployment-specific PV layout within these payloads. Figure~\ref{fig:wireshark} provides a concrete example using a SWaT EtherNet/IP/CIP packet whose deployment-specific PV layout is not resolved by the available dissector. Wireshark parses the standard protocol fields but exposes the PV-carrying payload only as command-specific data, whereas PVParser further decodes the corresponding individual PVs. PVParser therefore complements existing dissectors by recovering the missing deployment-specific PV-decoding layer rather than replacing standard protocol parsing.

Recovering this layer remains challenging because PV-carrying payloads are often long and deployment-specific, and their internal structures are often unavailable. A single payload may pack multiple PVs with non-PV fields such as padding, delimiters, or checksums, resulting in ambiguous field boundaries. As illustrated in Figure~\ref{fig:fig2.2}, even a small portion of a water treatment process contains diverse PVs with different physical meanings and value ranges. These PVs may be placed and encoded differently across industrial deployments because their layouts depend on plant settings or controller configurations. Such layouts may be documented only in deployment-specific engineering artifacts, which can be unavailable, incomplete, or outdated. Given these factors, recovering PVs further requires inferring PV fields within these payloads and associating them with PV semantics.

\subsection{PV Recovery Setting and Scope}
\label{sec:recovery}

This section describes the setting in which PVParser operates and the scope of PV recovery considered in this work.

\subsubsection{PV Recovery Setting}
PVParser targets legacy or vendor-maintained industrial CPS deployments where passively captured industrial network traffic and historian records are available, but the deployment-specific artifacts needed to decode PV-carrying payloads are incomplete, outdated, or unavailable. When complete and current artifacts are available, implementing and maintaining a deployment-specific dissector provides direct decoding. Reconfiguring the system to forward additional PVs or increase historian polling is also feasible when the running system can be safely modified and revalidated. PVParser is intended for settings where these conditions do not hold or where changes to the deployed system are undesirable, and automatically recovers the deployed PV layout from existing traffic and historian records.

PVParser operates on mirrored copies of communication already present on the industrial network. It decodes the PV-carrying payloads at the monitoring side and supplies the recovered PVs directly to downstream security monitoring. These PVs are not transmitted back to controllers or routed through the HMI or historian polling path. Therefore, PVParser adds monitoring-side capture and processing but does not generate additional controller requests or responses on the industrial network. In contrast, increasing historian polling introduces additional requests, responses, and controller processing on the industrial network.

\subsubsection{PV Recovery Scope}
Under this setting, this paper addresses the problem of recovering process variables from raw industrial network traffic to improve process visibility in industrial CPSs. As illustrated in Figure~\ref{fig:fig2.1}, given historian records and network traffic captured at the supervisory--control boundary, the goal is to recover a decoding specification for PV-carrying payloads. This specification describes how to identify PV-carrying payloads in network traffic, decode byte fields in these payloads into PV values, and associate the decoded fields with PV semantics. We assume that a basic protocol control field identifying the message type or operation is available, such as a Modbus function code or an EtherNet/IP/CIP service. Such fields can be obtained directly from standard protocol dissectors or through existing PRE techniques~\cite{qin2024reverse}.

We focus on recurrent and layout-stable PV-carrying payloads in plaintext, byte-oriented industrial protocols. Recurrent means that PV-carrying payloads appear repeatedly within the observation window, providing sufficient samples for recovery. Layout-stable means that PV-relevant fields appear in consistent byte-level positions and formats across payload instances. We primarily target periodic polling communication, which provides a reliable basis for observing recurrent PV-carrying payloads in industrial CPSs. Non-strictly-periodic payloads are within scope when they are repeatedly observable and share a stable payload layout. Under these conditions, PVParser applies to 13 of the 18 widely used ICS protocols reported by Censys~\cite{censys2024}, including Modbus/TCP, EtherNet/IP/CIP, S7comm, and FINS. Encrypted payloads, changed-only or event-driven messages with variable PV lists, and communication without sufficient recurrence or layout stability are outside the scope of this work.

\begin{figure}[!t]
	\centering
        \includegraphics[width=0.9\columnwidth]{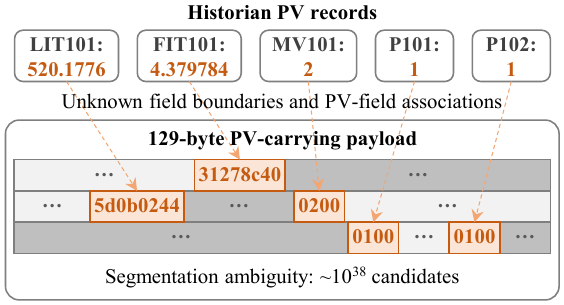}
	\caption{Example of PV field inference for a 129-byte PV-carrying payload. Even with historian PV records as references, field boundaries and PV--field associations remain unknown, resulting in a large candidate segmentation space.}
	\label{fig:mapping}
\end{figure}

\subsection{Technical Challenges}
Realizing the above visibility-recovery goal in industrial CPSs requires overcoming two core technical challenges: localizing PV-carrying payloads from heterogeneous industrial traffic and inferring PV fields within long deployment-specific payloads. We detail these challenges below.

\subsubsection{Challenge 1: Localizing PV-carrying Payloads in Heterogeneous Industrial Traffic}
Localizing PV-carrying payloads is challenging because they are embedded in mixed industrial traffic. In real deployments, PV-carrying messages are interleaved with other runtime messages, and messages with similar packet-level features, such as sizes, directions, or control fields, may correspond to different communication functions. Therefore, structural or statistical traffic characteristics alone are often insufficient to reliably distinguish PV-carrying communication from other industrial traffic.

\paragraph{Key Insight.}
PV-carrying communication in industrial CPSs often follows recurrent communication patterns with stable payload layouts. Periodic polling is a particularly reliable and widely observed form of such recurrence~\cite{stouffer2011guide}. It exhibits stable timing, consistent request-response roles, and recurring packet-level features. Such periodic regularities provide strong cues for localizing PV-carrying payloads from heterogeneous industrial traffic.

\paragraph{Our Solution.}
\tool{} localizes PV-carrying payloads by detecting periodic polling patterns with stable request-response roles and recurring packet-level features. It uses the detected patterns to recover the PV transmission network and extract PV-carrying payloads from mixed industrial traffic.

\subsubsection{Challenge 2: Inferring PV Fields in Long Deployment-Specific Payloads}
PV field inference faces substantial segmentation ambiguity in long deployment-specific payloads. Even moderate-length payloads can induce a vast combinatorial search space; as illustrated in Figure~\ref{fig:mapping}, a 129-byte payload containing only a few PVs already yields on the order of $10^{38}$ candidate segmentations. In such a large search space, sequential heuristic inference can commit to incorrect early boundary decisions because it performs greedy local search over field boundaries. These early errors propagate through the remaining segmentation process, preventing reliable recovery of later PV fields.

\paragraph{Key Insight.}
PV field recovery in long deployment-specific payloads should evaluate candidate segmentations at the payload level. A locally plausible boundary is reliable only if the resulting segmentation still supports the recovery of later PV fields. Therefore, inference should preserve multiple candidate segmentations and favor those that remain plausible across the payload.

\paragraph{Our Solution.}
\tool{} adopts a search-based optimization strategy for PV field inference. It explores multiple candidate segmentations under a constrained search budget and evaluates them using local boundary confidence and global structural consistency. This design enables \tool{} to identify high-quality field segmentations without prematurely committing to boundary decisions.

\begin{figure*}[!t]
	\centering
	\includegraphics[width=0.9\textwidth]{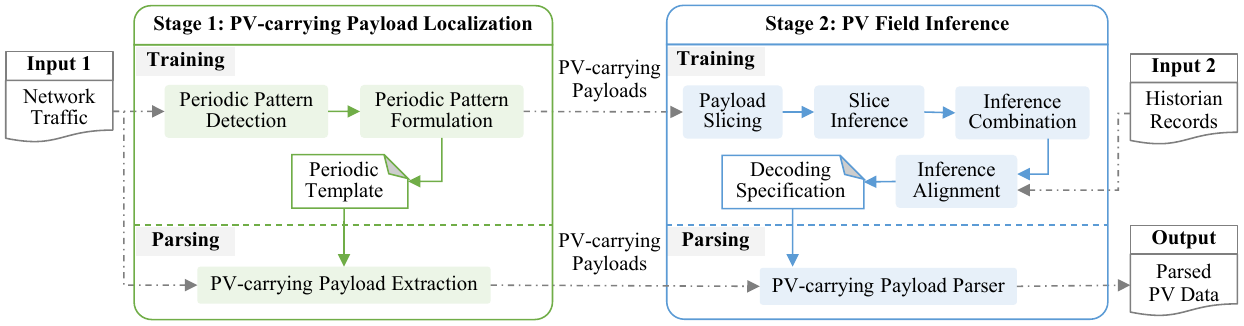}
	\caption{Architecture of \tool{}, which takes network traffic and historian records as inputs and outputs parsed PV data.}
	\label{fig:fig2.6}
\end{figure*}

\section{System Design}
To address the two challenges outlined above, we design \tool{} as a two-stage framework for recovering PVs from raw industrial network traffic, as illustrated in Figure~\ref{fig:fig2.6}. The first stage localizes PV-carrying payloads from heterogeneous industrial communication by exploiting periodic polling regularity between historians and PLCs. The second stage infers the internal structure of the extracted payloads through search-based segmentation, followed by semantic alignment with historian PV records. Together, these stages produce a complete decoding specification that enables direct parsing of PVs from network traffic in industrial CPSs.

\subsection{Problem Formulation}
We aim to recover PVs directly from raw industrial network traffic without relying on protocol documentation or intrusive system interaction. To support this goal, we consider captured network traffic $\mathcal{T}$ and historian PV records $\mathcal{H}$ as the only observable inputs. Our objective is to learn a mapping
\begin{equation}
\mathcal{M} : (\mathcal{T}, \mathcal{H}) \rightarrow \mathcal{S},
\;
\text{with } \mathcal{S}(\mathcal{T}) = \mathcal{V},
\end{equation}
\noindent
where $\mathcal{S}$ is a decoding specification that determines how to extract PV values $\mathcal{V}$ from traffic $\mathcal{T}$ at runtime.

This objective can be decomposed into two coupled subproblems. The first is to identify which messages in mixed industrial traffic actually carry PVs. The second is to recover the internal field structure of these PV-carrying payloads and associate the inferred fields with historian-observable PV semantics. The first subproblem reduces the search space by isolating relevant payloads from heterogeneous communication, while the second resolves the structural ambiguity within long deployment-specific payloads.

Based on this decomposition, \tool{} adopts a two-stage design. It first localizes PV-carrying payloads using periodic communication regularity, and then performs guided search over candidate payload segmentations to infer field boundaries, data types, endianness, and PV associations. This formulation allows \tool{} to avoid premature sequential decisions and instead recover a decoding specification through progressively refined structural and semantic inference.

\subsection{PV-Carrying Payload Localization}
To localize PV-carrying payloads from mixed industrial communication, \tool{} exploits the periodic polling relationship between historians and PLCs, an inherent characteristic of industrial control cycles~\cite{lin2019timing, villa2025icsquartz}. Such communication reflects regular system-state synchronization and thus provides a reliable cue for identifying PV-carrying transmissions. Based on this observation, \tool{} localizes PV-carrying payloads by detecting periodic communication patterns and formalizing them into reusable templates.

\subsubsection{Periodic Pattern Detection}
\label{sec:3.1.1}
Given raw network traffic $\mathcal{T}$, \tool{} detects stable periodic request-response cycles for each communicating pair and recovers the corresponding periodic patterns. It first filters and encodes packet streams into lightweight communication features, then identifies candidate periods through autocorrelation analysis and refines them via sliding-window matching. The resulting candidate periods are subsequently evaluated, and the best remaining one is selected after repeated periods are removed. Algorithm~\ref{alg:periodic_detection} summarizes this process.

\begin{algorithm}[ht]
\caption{Periodic Pattern Detection in Industrial Traffic}
\label{alg:periodic_detection}
\begin{algorithmic}[1]
\Require Raw network traffic $\mathcal{T}$, autocorrelation threshold $\theta_{acf}$
\Ensure Detected periodic pattern $p^*$

\State $\mathcal{T}' \leftarrow \textsc{FilterPackets}(\mathcal{T})$
\State $S \leftarrow \textsc{EncodePackets}(\mathcal{T}')$
\State $\mathcal{P}_c \leftarrow \textsc{DetectCandidatePeriods}(S, \theta_{acf})$

\For{each candidate period $p \in \mathcal{P}_c$}
    \State $\sigma_p \leftarrow \textsc{SlidingWindowMatch}(S, p)$
    \State $\mathbf{m}(p) \leftarrow \big(\mathrm{Cov}(\sigma_p),\ \mathrm{VarInt}(\sigma_p),\ \mathrm{VarDur}(\sigma_p)\big)$
\EndFor

\State $\mathcal{P}_c \leftarrow \textsc{RemoveRepeatedPeriods}(\mathcal{P}_c,\{\sigma_p\}_{p \in \mathcal{P}_c})$
\State $p^* \leftarrow \textsc{SelectBestPeriod}(\mathcal{P}_c, \mathbf{m})$
\State \Return $p^*$
\end{algorithmic}
\end{algorithm}

Specifically, \tool{} first filters out packets unrelated to industrial data transfer, such as handshake or retransmission packets, and encodes each remaining packet as
\begin{equation}
f_i = (dir_i, len_i, ctrl_i),
\end{equation}
where $dir_i$, $len_i$, and $ctrl_i$ denote packet direction, packet length, and application-level control code, respectively. These features preserve the communication structure most relevant to periodic-cycle detection while remaining agnostic to protocol documentation and payload semantics. The filtered traffic is thus transformed into a feature sequence
\begin{equation}
S = [f_1, f_2, \dots, f_n],
\end{equation}
as in Lines~1--2 of Algorithm~\ref{alg:periodic_detection}.

\tool{} then applies autocorrelation analysis to $S$ to identify dominant periodic components. Peaks exceeding a predefined threshold indicate candidate periods associated with recurring communication cycles:
\begin{equation}
\mathcal{P}_c = \{p \mid R_S(p) > \theta_{acf}\},
\end{equation}
where $R_S(p)$ is the autocorrelation score of sequence $S$ at lag $p$ (Line~3). Since autocorrelation only provides coarse period estimates and does not directly reveal the precise repeating segments, \tool{} further refines each candidate period using sliding-window matching. For each candidate period $p$, this step recovers a repeating segment sequence $\sigma_p$ under that period hypothesis, such that the direction, length, and control-code patterns remain consistent across cycles (Line~5).

\tool{} then evaluates the recovered sequence $\sigma_p$ using three metrics:
\begin{equation}
\mathbf{m}(p)=\big(\mathrm{Cov}(\sigma_p),\ \mathrm{VarInt}(\sigma_p),\ \mathrm{VarDur}(\sigma_p)\big),
\end{equation}
where $\mathrm{Cov}(\sigma_p)$ is its traffic coverage ratio, $\mathrm{VarInt}(\sigma_p)$ is the variance of inter-segment intervals, and $\mathrm{VarDur}(\sigma_p)$ is the variance of segment durations (Line~6). These metrics characterize whether the recovered sequence explains a substantial portion of the recurring traffic while remaining temporally stable.

To avoid selecting trivial multiples of a shorter underlying cycle, \tool{} removes candidate periods whose recovered segment patterns are simple repetitions of a previously identified shorter pattern (Line~8). The remaining candidate periods are then compared hierarchically by prioritizing larger coverage, then smaller interval variance, and finally smaller duration variance (Line~9). This ordering ensures that the selected pattern first captures the dominant recurring traffic and then favors temporal stability, with duration stability as a final refinement. \tool{} then groups the resulting patterns that share similar periods and originate from the same client host. The resulting groups characterize the PV transmission network between the supervisory and control levels~\cite{sheng2021cyber, ghazo2024andvi}.

\subsubsection{Periodic Pattern Formulation}
After periodic patterns are detected, \tool{} formalizes them into reusable templates for PV-carrying payload localization. For a communicating pair consisting of client $c$ and server $s$, the periodic template is represented as
\begin{equation}
P_{c,s} = \{ \text{``per\_len'': } T, \; \text{``per\_pat'': } [f_1, f_2, \dots, f_T] \},
\end{equation}
where ``per\_len'' denotes the period length and ``per\_pat'' denotes the ordered sequence of packet features within one cycle. Each $f_i$ reuses the packet-direction, length, and control-code encoding defined above. For example, for client $c = \text{192.168.1.200}$ and server $s = \text{192.168.1.20}$, a detected periodic pattern with $T = 2$ can be represented as
\begin{equation}
\begin{aligned}
P_{c,s} &=
\{ \text{``per\_len''}: 2, \; \text{``per\_pat''}: [f_1, f_2] \}, \\
f_1 &= \left(C ,\: 108 ,\: \text{0x0070:0x4C}\right), \\
f_2 &= \left(S ,\: 323 ,\: \text{0x0070:0xCC}\right),
\end{aligned}
\end{equation}
where \texttt{0x0070} denotes the EtherNet/IP Send Unit Data command, and \texttt{0x4C} and \texttt{0xCC} denote the CIP Read Tag request and response service codes, respectively.

These templates specify both the periodic communication structure and the response position within each cycle. Leveraging them, \tool{} matches traffic against the corresponding $P_{c,s}$ patterns and extracts the response payloads in matched cycles as PV-carrying payloads. The extracted payloads are then passed to the subsequent PV field inference stage, and the learned templates are reused for online payload parsing.

\subsection{PV Field Inference}
\label{sec:pv_field_inference}
Given the extracted PV-carrying payloads, \tool{} recovers their internal field structure and establishes their correspondence to historian-observable PV semantics through a staged search-based inference process. Rather than segmenting the entire payload in a single pass, \tool{} progressively refines the solution in four steps: payload slicing, slice-level field inference, inference combination, and semantic alignment. This staged design reduces the risk of premature boundary decisions in long deployment-specific payloads while preserving both structural coherence and semantic consistency in the final decoding result.

\begin{figure}[!t]
	\centering
        \includegraphics[width=0.9\columnwidth]{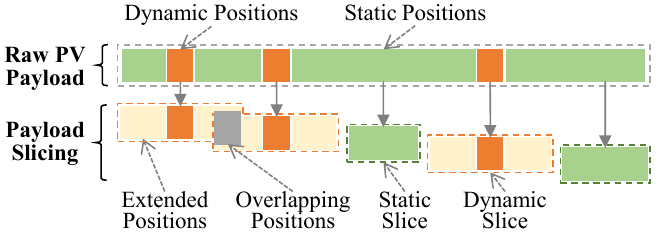}
    \caption{PV-carrying payload slicing, where dynamic slices are expanded from dynamic positions and static slices cover the remaining bytes.}
	\label{fig:fig2.7}
\end{figure}

\subsubsection{Payload Slicing}
To focus field inference on informative byte regions, \tool{} first partitions each PV-carrying payload into dynamic and static slices, as illustrated in Figure~\ref{fig:fig2.7}. Dynamic slices correspond to byte regions whose values vary across payload instances and are therefore more likely to encode process values, whereas static slices cover the remaining regions that stay unchanged over the observation window. This decomposition isolates informative byte regions for focused inference while preserving full payload coverage for subsequent reconstruction. Formally, let the PV-carrying payloads across periods be denoted as
\begin{equation}
\Phi = \{ \phi_1, \phi_2, \dots, \phi_n \}, \; |\phi_i| = L, \; \forall i,
\end{equation}
where $n$ is the number of periods and $L$ is the payload length.

\paragraph{Dynamic Slices.}
\tool{} first identifies dynamic byte positions whose values vary across the payload set:
\begin{equation}
\textit{DP} = \{ pos \mid \exists i,j: \phi_i[pos] \neq \phi_j[pos], \; \phi_i,\phi_j \in \Phi \}.
\end{equation}
Adjacent dynamic positions are merged and expanded by $\delta$ bytes on both sides to capture complete field boundaries and account for endianness, since the most-significant bytes of a multi-byte field may remain unchanged. Each dynamic slice is defined as
\begin{equation}
\textit{DS}_k = \{ pos \mid pos_k^{start} - \delta \le pos \le pos_k^{end} + \delta \},
\end{equation}
where $pos_k^{start}$ and $pos_k^{end}$ denote the first and last positions of the $k$-th dynamic region. The expansion size is set to $\delta = 7$ to cover the largest data types considered (UInt/Int64 and Float64). As shown in Figure~\ref{fig:fig2.7}, this expansion may introduce overlapping positions between adjacent dynamic slices, which are resolved later during inference combination.

\paragraph{Static Slices.}
The remaining payload positions are treated as static:
\begin{equation}
\textit{SP} = \{ pos \in \{0,1,\dots,L-1\} \mid pos \notin \textit{DS}_k \text{ for all } k \}.
\end{equation}
These positions are grouped into contiguous static slices,
\begin{equation}
\textit{SS}_r = \{ pos \in \textit{SP} \mid pos_r^{start} \le pos \le pos_r^{end} \},
\end{equation}
where $pos_r^{start}$ and $pos_r^{end}$ denote the first and last positions of the $r$-th static slice. Static slices complement dynamic ones to ensure complete payload coverage and preserve potentially meaningful fields whose values remain constant during the observation period.

\begin{figure}[!t]
	\centering
	\includegraphics[width=\columnwidth]{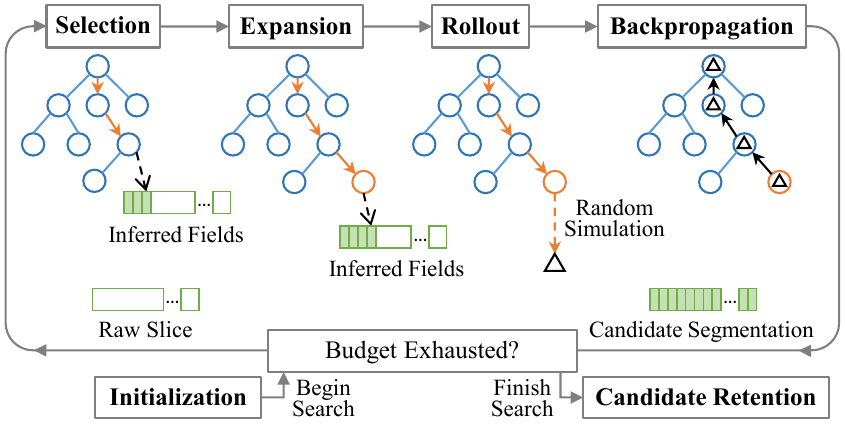}
	\caption{MCTS-based slice inference process, which iteratively explores the segmentation space through selection, expansion, rollout, and reward backpropagation.}
	\label{fig:mcts}
\end{figure}

\subsubsection{Slice Inference Engine}
\label{sec:slice_inference}
For each slice, \tool{} infers field boundaries through a tree-structured search over candidate segmentations. As illustrated in Figure~\ref{fig:mcts}, the search starts from an unsegmented slice and progressively expands partial field hypotheses through MCTS. This process combines informed node selection, heuristic expansion, rollout-based evaluation, and reward backpropagation to identify high-quality slice-level segmentations without committing to boundaries greedily. 
During the search, \tool{} jointly infers each field's offset, length, type, and endianness, which together form its decoding specification. The supported field types include Float32/Float64, signed and unsigned 8/16/32/64-bit integers, and ASCII/UTF-8 strings.

\paragraph{Initialization.}
For each slice, \tool{} initializes a root node representing the raw unsegmented byte sequence. The search then begins by expanding candidate field hypotheses from this root.

\paragraph{Selection.}
At each iteration, \tool{} selects an incomplete node using a UCB1-based criterion. This step balances exploration of new segmentation hypotheses and exploitation of promising ones by jointly considering backpropagated global segmentation rewards and local field plausibility. The UCB1 scoring function is defined as
\begin{equation}
\label{eq:ucb1_full}
\begin{aligned}
\text{UCB1}(n_i) &= R_{\text{exploit}}(n_i) + c\,\sqrt{\ln N_p / N_i}, \\
R_{\text{exploit}}(n_i) &= (1 - w_i)\,R_{\text{global}}(n_i) + w_i\,R_{\text{field}}(n_i), \\
R_{\text{global}}(n_i) &= \alpha\max(R_{\text{seg}}) + (1-\alpha)\operatorname{mean}(R_{\text{seg}}),
\end{aligned}
\end{equation}
where \(N_p\) and \(N_i\) denote the visit counts of the parent and current node, respectively, and \(c\) controls the exploration-exploitation balance. The depth-adaptive weight \(w_i\) increases with search depth. \(R_{\text{global}}(n_i)\) aggregates the backpropagated segmentation rewards \(R_{\text{seg}}\) from descendant rollouts, while \(R_{\text{field}}(n_i)\) represents the field-level confidence at node \(n_i\). 

At shallow depths, the UCB1 score emphasizes \(R_{\text{global}}\) to encourage broad exploration of segmentation trajectories; as depth increases, greater weight is placed on \(R_{\text{field}}\), guiding the search toward deeper exploitation and faster convergence to globally consistent segmentations.

\paragraph{Expansion.}
From the selected node, \tool{} infers the next candidate field using heuristic rules that assign a soft confidence score \(R_{\text{field}}\) to the proposed field type. These heuristics combine byte-level statistical cues, such as entropy profiles, with type-specific validity checks. For instance, float fields are evaluated through IEEE~754 conformance and plausible engineering ranges, integer fields through range utilization and numeric stability, and string fields through printable-character ratios and encoding validity. The resulting field hypothesis is appended as a child node, extending the partial segmentation. Heuristic details are given in Appendix~\ref{sec:appA}.

\paragraph{Rollout.}
To evaluate the feasibility of the partial segmentation, \tool{} stochastically completes the remaining field boundaries from the newly expanded child node. This rollout assesses whether the current boundary decision can be extended into a coherent slice-level segmentation. The resulting full segmentation is then scored by a structural reward \(R_{\text{seg}}\) that aggregates coverage completeness, boundary-confidence evidence, and entropy-pattern consistency.

\paragraph{Backpropagation.}
The structural reward is backpropagated along the visited path to update MCTS statistics, increasing the priority of segmentation decisions that are more likely to yield correct field boundaries.

\paragraph{Candidate Retention.}
The search iterates until the per-slice MCTS budget is exhausted. The budget is proportional to the slice length, allowing longer slices to receive more search iterations. To account for byte-order ambiguity, each slice is inferred under both little-endian and big-endian assumptions. Under each assumption, \tool{} retains the top-$k$ slice-level segmentation candidates $\mathcal{Z}$ ranked by structural reward.

\begin{figure}[!t]
	\centering
    \includegraphics[width=0.9\columnwidth]{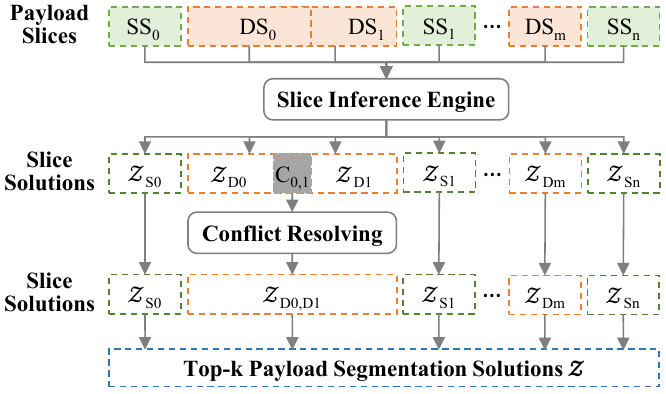}
    \caption{Inference combination integrating dynamic and static slice results into a complete and consistent PV-carrying payload segmentation.}
	\label{fig:fig2.8}
\end{figure}

\subsubsection{Inference Combination}
To obtain coherent segmentations for the entire PV-carrying payload, \tool{} integrates the slice-level candidates inferred from dynamic and static slices, as illustrated in Figure~\ref{fig:fig2.8}. This stage lifts inference from the slice level to the payload level by resolving overlapping boundaries, reconciling local field decisions, and retaining only the most plausible full-payload segmentations.

The combination proceeds in two steps. First, for regions shared by multiple dynamic slices, \tool{} performs conflict resolving to reconcile overlapping slice solutions and eliminate incompatible boundary assignments. To do so, it re-invokes the \emph{Slice Inference Engine} on the overlapping regions and re-evaluates the resulting candidates under slice-combination constraints. Second, the reconciled dynamic slices are merged with static slices to form complete payload-level segmentations with non-overlapping coverage and consistent endianness. To avoid combinatorial explosion as the number of slices grows, \tool{} employs beam search~\cite{deutschmann2024conformal} to retain only the top candidates during hierarchical merging. The final output of this stage is a top-$k$ set of payload segmentation solutions \(\mathcal{Z}\), which are then passed to the alignment stage.

\subsubsection{Inference Alignment}
\label{sec:inference_alignment}
After payload-level segmentation candidates are obtained, \tool{} establishes the final correspondence between inferred payload fields and historian-observable PVs. This stage resolves the semantics of the recovered field structures by aligning inferred field groups with PV groups derived from historian records. Based on the resulting alignment, \tool{} produces the decoding specification used for PV parsing.

\paragraph{Group Definition.}
The inferred fields after the combination stage are organized by historian--PLC communication pairs and their associated segmentation candidates:
\begin{equation}
\mathcal{G}_f = \{\, G_{f}^{(c, s, \zeta_k)} \mid c \in C,\, s \in S,\, \zeta_k \in \mathcal{Z} \,\},
\end{equation}
where \(c\) and \(s\) denote the historian and PLC IP addresses, respectively, and \(\zeta_k\) is the \(k\)-th segmentation candidate in \(\mathcal{Z}\).

Historian PVs are grouped by tag pattern, where tags sharing the same process unit identifier are assigned to the same PV group. Let \(\mathcal{I}\) denote the set of PV-group identifiers. The resulting historian PV groups are defined as
\begin{equation}
\mathcal{G}_h = \{\, G_{h}^{(i)} \mid i \in \mathcal{I} \,\}.
\end{equation}

\paragraph{Alignment Process.}
For each candidate group pair $(G^{(c,s,\zeta_k)}_f,G^{(i)}_h)$,
\tool{} aligns inferred field sequences with historian PV sequences by maximizing a normalized multi-metric alignment score. This score integrates three complementary signals: DTW-based trajectory similarity, distributional similarity derived from KL divergence, and range consistency. DTW captures temporal shape agreement under sampling jitter and timing offsets, KL-based similarity penalizes candidates with inconsistent value distributions, and range consistency penalizes decoded values outside plausible physical ranges. The three normalized signals are combined into a single alignment score using fixed weights, and the candidate PV--field mapping with the highest score is selected.

This alignment yields a mapping between PV semantics and inferred fields:
\begin{equation}
\label{eq:mapping}
\begin{aligned}
\mathcal{M}_{PV}
  &= \{\, \text{PV}_i \rightarrow \text{FieldSpec}_i \,\}, \\
\text{PV}_i
  &= \{\textit{group\_id},\, \textit{pv\_name}\}, \\
\text{FieldSpec}_i
  &= \{\textit{sess\_key},\, \textit{sol\_idx},\, \textit{field\_spec}\},
\end{aligned}
\end{equation}
where \textit{group\_id} denotes the PV group index, \textit{pv\_name} is the historian PV tag, \textit{sess\_key} = \((c,s)\) specifies the historian--PLC pair, \textit{sol\_idx} indexes the selected segmentation candidate, and \textit{field\_spec} records the inferred field offset, length, type, endianness, and confidence.

\paragraph{PV-carrying Payload Parser.}
Once structural inference and semantic alignment are both resolved, \tool{} synthesizes a complete decoding specification for PV-carrying payload parsing. The recovered field structures and PV--field mappings are exported in a structured JSON format, enabling parsing, interoperability, and reuse across programming languages and platforms. An example of the generated decoding entry is provided in Appendix~\ref{sec:appB}.

\section{Experimental Evaluation}

This section evaluates \tool{} in two layers. The first layer examines technical correctness and robustness under verifiable reference data. Historian-observable PVs provide reliable ground truth for PV field recovery, enabling consistent comparison with prior PRE methods. The second layer examines whether the recovered visibility goes beyond historian records and benefits security monitoring.

Guided by this design, we answer five evaluation questions: 
(1) whether \tool{} can reliably localize PV-carrying communication from mixed industrial traffic; 
(2) whether it can accurately recover PV fields and semantics compared with prior PRE methods; 
(3) whether its key design components are necessary for accurate PV recovery; 
(4) whether \tool{} remains stable under variations in key parameters and how parameter-induced errors propagate through the recovery pipeline; and 
(5) whether \tool{} can recover control-level PVs beyond historian records and improve downstream attack detection. 
Beyond these questions, Appendix~\ref{sec:window} analyzes how observation-window length affects PV recovery performance, and Appendix~\ref{sec:efficiency} evaluates the execution-time efficiency of \tool{}.

\paragraph{Industrial CPS Benchmarks.}
We use three widely adopted CPS datasets for evaluation: SWaT~\cite{goh2016dataset}, WADI~\cite{ahmed2017wadi}, and SCADA Network (SCADA-N)~\cite{lemay2016providing}. 
SWaT models a six-stage water treatment plant with continuous closed-loop control and provides both network traffic and historian records, enabling evaluation of PV field recovery. 
WADI represents a large-scale water distribution system with interconnected storage and distribution stages and likewise provides network traffic and historian records. 
SCADA-N provides representative SCADA communication traces generated in a controlled laboratory environment, but does not include historian records, making it suitable only for evaluating periodic pattern detection.

\paragraph{State-of-the-Art Baselines.}
Recovering PVs directly from raw industrial traffic has not been addressed in prior work. 
Therefore, there is no directly comparable baseline for the full \tool{} pipeline. 
Instead, we select six representative PRE methods as baselines to benchmark the field-inference and semantic-recovery capabilities of \tool{}: NetPlier~\cite{ye2021netplier}, BinaryInferno~\cite{chandler2023binaryinferno}, CAN-D~\cite{verma2021can}, ByCAN~\cite{lin2024bycan}, DP-Reverser~\cite{yu2022towards}, and SeMiner~\cite{cai2022seminer}.

These baselines represent different technical families. 
NetPlier infers message formats by probabilistically aligning message instances and learning field structures and types. 
BinaryInferno identifies field boundaries and data types using a semantic-guided ensemble of binary-analysis techniques. 
CAN-D follows a modular signal-decoding pipeline that identifies field offsets, encodings, and physical meanings in automotive communication traces, while ByCAN enhances this design through byte- and bit-level clustering for more precise slicing and labeling. 
DP-Reverser focuses on variable-field alignment by fitting relationships between raw communication values and observed sensor or actuator signals. 
SeMiner leverages side information from industrial processes to extract protocol semantics and associate fields with physical meanings.

\begin{table}[t]
\centering
\caption{Coverage of baselines across PV recovery stages.}
\label{tab:alignment}
\small
\begin{threeparttable}
\begin{tabular}{
>{\centering\arraybackslash}p{2.1cm}
>{\centering\arraybackslash}p{1.5cm}
>{\centering\arraybackslash}p{1.5cm}
>{\centering\arraybackslash}p{1.5cm}
}
\toprule
\textbf{Method} & \textbf{Localization} & \textbf{Inference} & \textbf{Alignment} \\
\midrule
NetPlier      & $\times$   & $\checkmark$ & $\times$   \\
BinaryInferno & $\times$   & $\checkmark$ & $\times$   \\
CAN-D         & $\times$   & $\checkmark$ & $\checkmark$   \\
ByCAN         & $\times$   & $\checkmark$ & $\checkmark$ \\
DP-Reverser   & $\times$   & $\times$ & $\checkmark$ \\
SeMiner       & $\times$   & $\checkmark$ & $\checkmark$ \\
\tool{}       & $\checkmark$   & $\checkmark$ & $\checkmark$ \\
\bottomrule
\end{tabular}
\begin{tablenotes}[para]
Note: Localization, Inference, and Alignment denote PV-carrying payload localization, PV field inference, and field semantic alignment.
\end{tablenotes}
\end{threeparttable}
\end{table}

\paragraph{Baseline Adaptation.}
Following prior PRE evaluation practice~\cite{buscemi2023survey, katcher2025investigation}, PV recovery can be viewed as a three-stage process: 
(1) locating PV-carrying payloads within mixed traffic, 
(2) inferring field boundaries inside these payloads, and 
(3) aligning recovered fields with process variables. 
The selected baselines cover different subsets of these stages, as summarized in Table~\ref{tab:alignment}. 
We therefore adapt each method only within the scope of its original capabilities.

Methods that infer semantics but do not isolate PV-carrying payloads (e.g., CAN-D, ByCAN, and SeMiner) are evaluated on the payloads localized by \tool{}.
Methods that infer field boundaries but not semantics (e.g., NetPlier and BinaryInferno) output candidate field segments, which we align to historian records using the \textit{Inference Alignment} mechanism in Section~\ref{sec:inference_alignment}. 
DP-Reverser, which assumes candidate field boundaries and performs variable-field alignment, is applied over sliding byte windows to generate candidate fields; for each candidate, we run its regression-based mapping and retain the best-fitting correspondence. 
These capability-preserving adaptations do not extend the native inference logic of the baselines, but enable a fair comparison under the same experimental setup and evaluation criteria. 
Because these baselines target different PRE tasks and use different datasets and metrics, the scores reported in their original studies are not directly comparable with those obtained under our PV-recovery setting.

\paragraph{Implementation and Environment.}
We implemented \tool{} in Python. The key parameters of \tool{} are fixed across datasets and communication paths, with their settings reported in Appendix~\ref{app:parameters}. We conducted all experiments on a Linux workstation with 32 CPU cores (Threadripper 2990WX, 3.0 GHz) and 128 GB of memory.

\begin{table}[t]
\centering
\caption{Periodic-pattern detection and recovered historian--PLC PV transmission paths.}
\label{tab:periodic_summary}
\small
\begin{tabular}{
>{\centering\arraybackslash}p{1.2cm}
>{\centering\arraybackslash}p{1.8cm}
>{\centering\arraybackslash}p{2.0cm}
>{\centering\arraybackslash}p{1.2cm}
}
\toprule
\textbf{Dataset} & \textbf{Validation} & \textbf{Result} & \textbf{PV paths} \\
\midrule
SCADA-N & Ground truth & 6/6 correct; avg.\ 10.12\,s (GT: 10\,s) & 6 \\
SWaT & Cross-segment consistency & 100\% consistent across 6 PCAPs & 6 \\
WADI & Cross-segment consistency & 100\% consistent across 6 PCAPs & 3 \\
\bottomrule
\end{tabular}
\end{table}

\begin{table}[t]
\centering
\caption{PV-carrying payload lengths (bytes) across PLCs.}
\label{tab:pv-length}
\small
\begin{tabular}{
>{\centering\arraybackslash}p{1.0cm}
>{\centering\arraybackslash}p{0.7cm}
>{\centering\arraybackslash}p{0.7cm}
>{\centering\arraybackslash}p{0.7cm}
>{\centering\arraybackslash}p{0.7cm}
>{\centering\arraybackslash}p{0.7cm}
>{\centering\arraybackslash}p{0.7cm}
}
\toprule
\textbf{Dataset} & \textbf{PLC\_1} & \textbf{PLC\_2} & \textbf{PLC\_3} & \textbf{PLC\_4} & \textbf{PLC\_5} & \textbf{PLC\_6} \\
\midrule
SWaT & 129 & 218 & 154 & 154 & 336 & 90 \\
WADI & 504 & 1469 & 401 & -- & -- & -- \\
\bottomrule
\end{tabular}
\end{table}

\subsection{Evaluation of Periodic Pattern Detection}
\label{sec:4.1}
To answer the first evaluation question, we evaluate \tool{}'s ability to recover periodic polling structures and the corresponding PV transmission network from mixed industrial traffic. 
For polling-cycle recovery, SCADA-N provides explicit ground truth, allowing direct period validation in its 6-controller setting. 
For SWaT and WADI, no ground-truth polling periods are available. We therefore evaluate correctness through cross-segment consistency. Specifically, we split 30 minutes of traffic into six 5-minute PCAP files and examine whether \tool{} recovers the same periodic communication patterns from each segment.
For transmission-network recovery, all three datasets document the relevant communication topology, enabling direct verification of the recovered network.

Table~\ref{tab:periodic_summary} summarizes the periodic-pattern detection and PV-path recovery results. In SCADA-N, \tool{} correctly recovers all six polling channels and estimates an average polling interval of 10.12\,s, closely matching the 10-second ground truth. This small deviation is likely due to slight timing variations in the captured traffic.
In SWaT and WADI, all detected periodic sessions remain fully consistent across independent PCAP segments, indicating stable recovery under realistic industrial conditions. 
In addition, the recovered polling paths match the documented system architectures in both path count and communication-pair identity.

SWaT further illustrates the selectivity of \tool{} in distinguishing PV-carrying communication from generic periodic traffic. Although 17 periodic device pairs are observed in total, only the historian server (192.168.1.200) consistently initiates stable polling toward six PLCs, revealing the actual PV transmission backbone rather than auxiliary periodic paths. 
Further analysis shows that PLC\_1--PLC\_5 operate at approximately 0.297\,s polling intervals, whereas PLC\_6 operates at a faster 0.047\,s polling interval, likely due to tighter real-time monitoring requirements.

Since SWaT and WADI do not provide ground-truth PV-carrying payload positions, we assess extraction correctness indirectly via the downstream field-inference results and summarize the resulting payload characteristics.
As shown in Table~\ref{tab:pv-length}, payload lengths vary substantially across PLCs and datasets, from 90 bytes in SWaT to 1,469 bytes in WADI. This heterogeneity directly affects downstream field inference because longer payloads induce a larger segmentation search space and greater structural ambiguity.

These results show that \tool{} reliably recovers periodic polling structures and PV transmission networks from mixed industrial traffic, providing a sound basis for localizing PV-carrying payloads before field inference.

\begin{table*}[t]
\centering
\caption{Performance comparison on SWaT and WADI datasets (35-minute training window).}
\label{tab:performance_35min}
\small
\setlength{\tabcolsep}{6pt}
\begin{tabular}{lcccccccc}
\toprule
\multirow{2}{*}{\textbf{Method}} &
\multicolumn{4}{c}{\textbf{SWaT}} &
\multicolumn{4}{c}{\textbf{WADI}} \\
\cmidrule(lr){2-5} \cmidrule(lr){6-9}
 & \textbf{Precision} & \textbf{Recall} & \textbf{F1-score} & \textbf{Accuracy} 
 & \textbf{Precision} & \textbf{Recall} & \textbf{F1-score} & \textbf{Accuracy} \\
\midrule
NetPlier      &   0.8571   &   0.5455   &  0.6667    &  0.5000    &  0.5000    &  0.1667    &  0.2500    & 0.1429     \\

BinaryInferno &   0.9231   &  0.3429    &  0.5000    &  0.3333    &  0.8889   & 0.2963     & 0.4444     &  0.2857    \\

CAN-D         &   0.5758   &  0.8636    &  0.6909    &  0.5278    &  0.5455    &  0.6667    &  0.6000    &  0.4286    \\

ByCAN         &   0.5000   &   0.8000   &  0.6154    &  0.4444    &  0.5385    &  0.8750    &  0.6667    & 0.5000     \\

DP-Reverser   &   0.6250   &  0.8333    &  0.7143    &  0.5556    &  0.8667   &  0.5000    &  0.6341    &  0.4643    \\

SeMiner       &   0.8750   &   0.6364   &   0.7368   &  0.5833    &  0.5882    & 0.4762   &  0.5263  &  0.3571  \\

\tool{}      &   \textbf{1.0000}   &  \textbf{0.9722}    &  \textbf{0.9859}    &  \textbf{0.9722}    &   \textbf{1.0000}   &  \textbf{0.9286}    &  \textbf{0.9630}    &  \textbf{0.9286}    \\
\bottomrule
\end{tabular}
\end{table*}

\subsection{Evaluation of PV Field Inference}
\label{sec:4.2}
To answer the second evaluation question, we benchmark \tool{} against six state-of-the-art baselines on the SWaT and WADI datasets to evaluate its accuracy in recovering PV fields and semantics. Following standard practice in PRE~\cite{yu2022towards, lin2024bycan}, we retain only dynamic historian PVs that exhibit temporal variation during operation, while excluding constant PVs that provide limited evidence for inference. Accordingly, 36 PVs (26 sensors and 10 actuators) from SWaT and 28 PVs (24 sensors and 4 actuators) from WADI are used in our evaluation. For all methods, PV--field mappings are derived from a 35-minute training window and validated on a non-contiguous 15-minute segment collected two hours later to assess generalization across operational states. Each method is applied to the PV-carrying payloads localized in Section~\ref{sec:4.1}.

A recovered PV--field mapping is counted as correct only if the inferred offset, length, type, and endianness jointly produce decoded values that match the corresponding historian PV records under the criterion defined below.
Since no public ground truth exists for PV--field correspondences, we follow the manual verification practice widely adopted in prior PRE studies~\cite{verma2021can, yu2022towards, lin2024bycan}. For each historian record $v_k^{h}(t_i)$, we examine payloads within a window 
$[t_i-\alpha,\, t_i+\alpha]$, where $\alpha=2$ seconds, to compensate for time offsets~\cite{cai2022seminer}. For each payload in this window, we decode the inferred field value $v_k^{f}(t_j)$ and compare it with the historian value $v_k^{h}(t_i)$. 
For continuous PVs, decoded values are counted as matched if their absolute error from the corresponding historian value is below $10^{-3}$, whereas discrete actuator states require exact equality. 
A PV field is regarded as correct if more than $95\%$ of its historian records are matched. 
This criterion accommodates occasional packet loss and transient value noise while maintaining a strict field-level correctness requirement.

We report four standard metrics used in prior PRE evaluation~\cite{chandler2023binaryinferno, sheng2026reverse}: \textit{Precision}, \textit{Recall}, \textit{F1-score}, and \textit{Accuracy}. Here, TP denotes correctly inferred PVs, FP denotes incorrectly inferred PVs, and FN denotes missed PVs; TN is not applicable to this task. The metrics are computed as follows:

\begin{equation}
\label{eq:metrics}
\begin{aligned}
\mathit{Precision} &=
\frac{\text{Correctly Inferred PVs (TP)}}{\text{Inferred PVs (TP + FP)}}, \\[4pt]
\mathit{Recall} &=
\frac{\text{Correctly Inferred PVs (TP)}}{\text{Expected PVs (TP + FN)}}, \\[4pt]
\mathit{F1\text{-}score} &=
\frac{2 \times \mathit{Precision} \times \mathit{Recall}}
{\mathit{Precision} + \mathit{Recall}}, \\[4pt]
\mathit{Accuracy} &=
\frac{\text{Correctly Inferred PVs (TP)}}{\text{Total PVs (TP + FP + FN)}}.
\end{aligned}
\end{equation}

As shown in Table~\ref{tab:performance_35min}, \tool{} consistently achieves the highest precision, recall, F1-score, and accuracy on both datasets. The advantage is especially notable on WADI, whose much longer payloads (up to 1,469 bytes, compared with 336 bytes in SWaT) substantially enlarge the inference search space and increase structural heterogeneity. Despite this increase in complexity, \tool{} maintains stable performance and a clear lead over all baselines, demonstrating strong robustness on long industrial payloads. 
The remaining recall loss on WADI arises from two PVs with weak temporal variation. Limited temporal variation makes their corresponding fields difficult to distinguish from non-PV fields during field inference, preventing recovery of the associated historian PVs.

We further examine constant PVs and the matching criterion. Among historian PVs that remain constant and can be evaluated unambiguously, \tool{} recovers 1 of 5 on SWaT and misses the single one on WADI. The missed constant PVs share the same limitation as the two WADI PVs discussed above. When the matching criterion is tightened from 95\% to 100\% on SWaT, \tool{} correctly recovers 24 of 36 dynamic PVs. The reduction mainly arises from occasional mismatches between packet-level values and historian samples because packet-level PV transmissions occur more frequently than historian samples are recorded. Such mismatches are tolerated by the 95\% criterion but not by the 100\% criterion.

The baselines exhibit substantially weaker and less balanced performance. NetPlier and BinaryInferno achieve relatively high precision on some settings but suffer from low recall, indicating that their fixed byte-pattern heuristics can identify only a limited subset of valid PV fields. CAN-D, ByCAN, DP-Reverser, and SeMiner recover more candidate fields, but their precision and overall accuracy remain low, suggesting difficulty in resolving ambiguous PV--field associations in real industrial processes. This weakness becomes more pronounced on WADI, where longer and more heterogeneous payloads amplify structural ambiguity and make sequential inference errors more costly.

These results are consistent with the design differences between \tool{} and the baselines. Most baselines rely primarily on local heuristics or sequential inference. In contrast, \tool{} uses a search-based inference framework that balances local confidence with global consistency through a modified MCTS guided by byte-pattern and process-aware heuristics. This design helps avoid early inference traps and better preserves globally coherent segmentations in long payloads.


\begin{figure}
    \centering
    \includegraphics[width=0.9\linewidth]{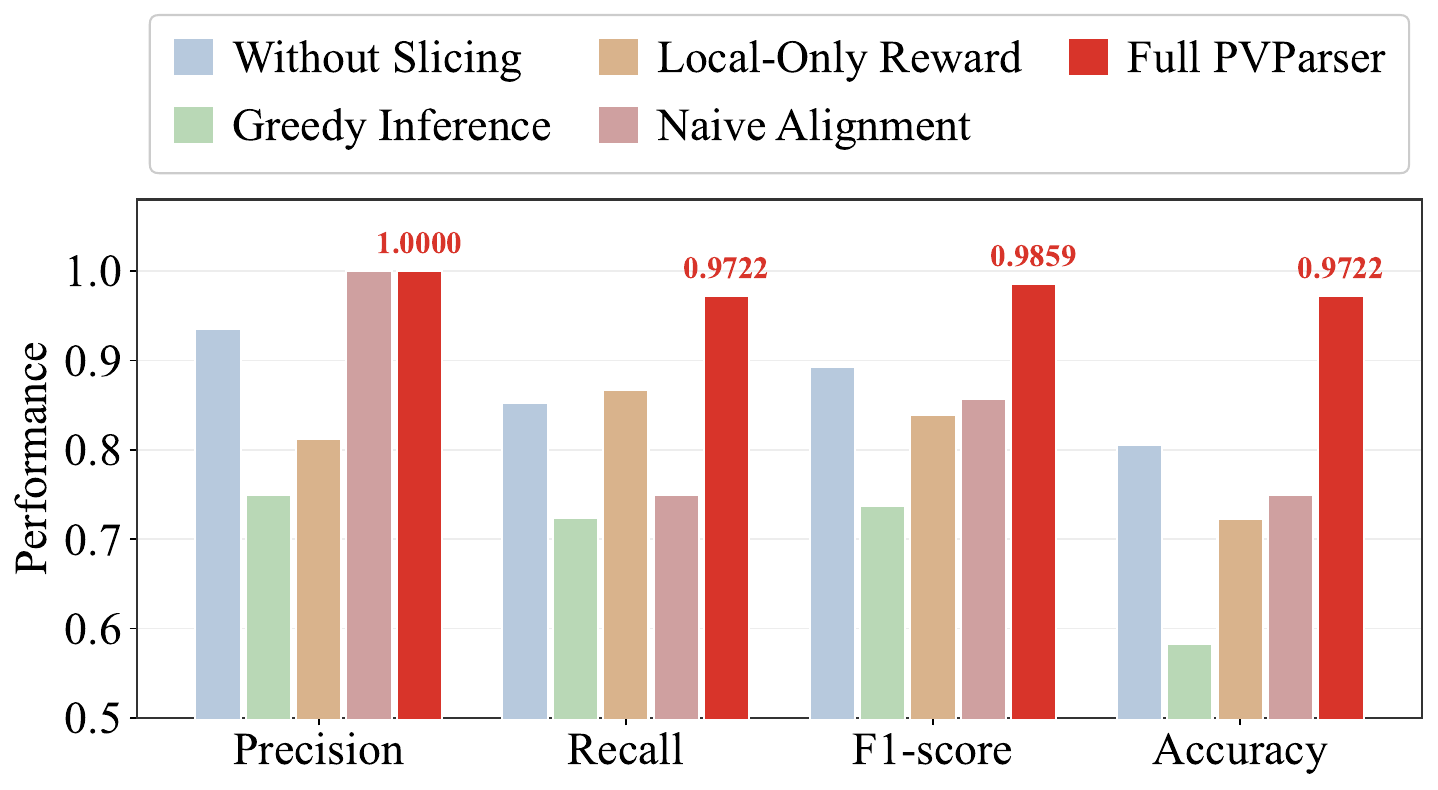}
    \caption{Ablation results of \tool{} on SWaT.}
    \label{fig:ablation}
\end{figure}

\subsection{Ablation Experiments}
\label{sec:ablation}

To answer the third evaluation question, we conduct an ablation study of \tool{}'s core components on SWaT. We compare the full \tool{} with four variants: \textbf{(1) Without Slicing}, which removes payload slicing and performs inference over the entire payload; \textbf{(2) Greedy Inference}, which replaces the modified MCTS with sequential greedy inference; \textbf{(3) Local-Only Reward}, which removes the global structural reward and relies only on local field plausibility during search; and \textbf{(4) Naive Alignment}, which replaces historian-based multi-metric alignment with DTW-based matching only~\cite{cai2022seminer, lin2024bycan}. All other settings remain unchanged.

As shown in Figure~\ref{fig:ablation}, all four variants underperform the full \tool{}. Removing payload slicing reduces F1-score from 0.9859 to 0.8923 and accuracy from 0.9722 to 0.8056. This indicates that slicing helps isolate informative byte regions and limit error propagation in long payloads.
Replacing modified MCTS with greedy inference causes the largest degradation across all metrics, with F1-score dropping to 0.7368 and accuracy to 0.5833. This shows that search-based optimization is critical for reducing segmentation errors. 
Using only local rewards also leads to a clear decline, indicating that local plausibility alone is insufficient and that explicitly modeling global segmentation consistency is important for identifying coherent payload structures.
Finally, Naive Alignment achieves precision of 1.0000 but with lower recall and accuracy. This result shows that accurate semantic recovery requires not only plausible field boundaries, but also robust alignment criteria.

Overall, the ablation results show that \tool{}'s performance gains are not attributable to any single component, but arise from the combined contribution of its key design choices.

\begin{figure}[!t]
    \centering
    \subfloat[Autocorrelation threshold.]{
        \includegraphics[width=0.52\columnwidth]{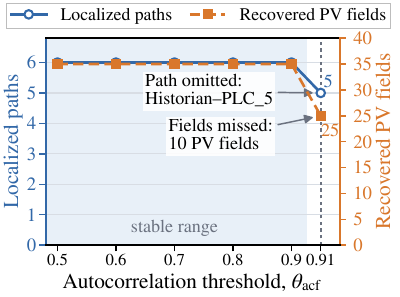}
    }
    \subfloat[Slice expansion size.]{
        \includegraphics[width=0.45\columnwidth]{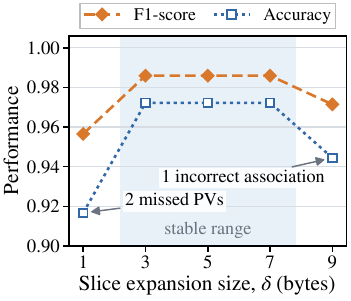}
    }
    \caption{Sensitivity of PVParser to the autocorrelation threshold
$\theta_{\mathrm{acf}}$ and slice expansion size $\delta$ on SWaT.}
    \label{fig:sensitivity}
\end{figure}

\subsection{Parameter Sensitivity and Error Propagation}
\label{sec:sensitivity}

To answer the fourth evaluation question, we evaluate the sensitivity of \tool{} to the autocorrelation threshold $\theta_{\mathrm{acf}}$ and slice expansion size $\delta$ on SWaT. We vary one parameter at a time while keeping all other settings unchanged. Figure~\ref{fig:sensitivity} reports the resulting localization and PV field recovery performance.

\paragraph{Autocorrelation Threshold.}
When $\theta_{\mathrm{acf}}$ varies from 0.5 to 0.9, \tool{} consistently localizes all 6 PV transmission paths and correctly recovers the same 35 PV fields. At $\theta_{\mathrm{acf}}=0.91$, the stricter threshold excludes the Historian--PLC\_5 path, reducing the numbers of localized paths and correctly recovered PV fields from 6 and 35 to 5 and 25, respectively. This localization error propagates through the pipeline because the PV-carrying payloads on the omitted path are not passed to field inference, causing all 10 PV fields carried on this path to be missed.

\paragraph{Slice Expansion Size.}
When $\delta\in\{3,5,7\}$, the F1-score and accuracy remain unchanged at 0.9859 and 0.9722, respectively. At $\delta=1$, the expanded slices provide insufficient boundary context for some four-byte continuous fields. Consequently, PIT501 and PIT503 are no longer retained as valid Float32 candidates and are missed during semantic alignment. At $\delta=9$, the enlarged overlapping regions introduce additional segmentation ambiguity during local re-inference, resulting in one incorrect PV association. These results show that an expansion size that is too small may miss fields near slice boundaries, whereas an excessively large size may introduce ambiguous candidates.

Overall, \tool{} remains stable across moderate parameter ranges. The observed errors follow the two-stage recovery pipeline: localization errors prevent all downstream field recovery on an omitted transmission path, whereas slice-expansion errors affect individual field inference and PV association.

\subsection{Visibility Extension and Security Utility}
To answer the fifth evaluation question, we evaluate \tool{} on SWaT through two connected analyses: recovering control-level PVs beyond historian records and assessing downstream attack detection using the expanded process view.

\begin{table}[t]
\centering
\caption{Recovered PVs from recurrent communication paths between PLC\_2 and peer PLCs in SWaT.}
\label{tab:pv_recovery}
\small
\begin{threeparttable}
\begin{tabular}{l c c c c c c}
\toprule
\textbf{Peer PLC} & \textbf{Ref.} & \textbf{Rec.} &
\textbf{Prec.} & \textbf{Recall} & \textbf{F1-score} & \textbf{Acc.} \\
\midrule
PLC\_1 & 2 & 2 & 1.0000 & 1.0000 & 1.0000 & 1.0000 \\
PLC\_3 & 5 & 4 & 1.0000 & 0.8000 & 0.8889 & 0.8000 \\
PLC\_4 & 1 & 1 & 1.0000 & 1.0000 & 1.0000 & 1.0000 \\
PLC\_5 & 4 & 4 & 1.0000 & 1.0000 & 1.0000 & 1.0000 \\
PLC\_6 & 1 & 1 & 1.0000 & 1.0000 & 1.0000 & 1.0000 \\
\midrule
\textbf{Total} & \textbf{13} & \textbf{12} &
\textbf{1.0000} & \textbf{0.9231} & \textbf{0.9600} & \textbf{0.9231} \\
\bottomrule
\end{tabular}
\begin{tablenotes}[para]
Note: Ref. and Rec. denote the numbers of reference and recovered PVs; Prec. and Acc. denote precision and accuracy.
\end{tablenotes}
\end{threeparttable}
\end{table}

\subsubsection{Recovering Control-Level PVs}
We evaluate whether \tool{} can recover PV values carried in control-level communication but absent from historian records. 
Since SWaT does not provide official annotations for these fields, two domain experts with PRE and industrial CPS experience independently inspected recurrent PLC-to-PLC communication paths involving PLC\_2. They identified candidate PVs using SWaT process documentation, repeated traffic patterns, and the physical plausibility of candidate value sequences. We retained only candidates identified by both experts, forming a conservative high-confidence reference set rather than an exhaustive inventory of all PLC-to-PLC PVs. 
We then apply \tool{} to the traffic on these control-level communication paths and compare the recovered PVs with the reference set.

As shown in Table~\ref{tab:pv_recovery}, the reference set contains 13 control-level PVs transmitted between PLC\_2 and other PLCs. \tool{} correctly recovers 12 of them. The only missed PV is \texttt{MV301} on the PLC\_2-to-PLC\_3 path, which remains static throughout the observation window and therefore provides insufficient temporal variation for field inference. Overall, \tool{} achieves 1.0000 precision, 0.9231 recall, 0.9600 F1-score, and 0.9231 accuracy. These results show that \tool{} can recover PV values from recurrent control-level communication, thereby extending process visibility beyond historian records. The recovered control-level PVs provide the expanded process view for downstream attack detection evaluation.

\begin{table*}[t]
\centering
\caption{Attack detection evaluation under four representative attack scenarios.}
\label{tab:security_eval}
\small
\setlength{\tabcolsep}{4pt}
\renewcommand{\arraystretch}{1.15}
\begin{threeparttable}
\begin{tabular}{lccc ccc ccc ccc}
\toprule
\multirow{2}{*}{\textbf{Method}} 
& \multicolumn{3}{c}{\textbf{S1}}
& \multicolumn{3}{c}{\textbf{S2}}
& \multicolumn{3}{c}{\textbf{S3}}
& \multicolumn{3}{c}{\textbf{S4}} \\
\cmidrule(lr){2-4}\cmidrule(lr){5-7}\cmidrule(lr){8-10}\cmidrule(lr){11-13}
& \textbf{F1-score} & \textbf{Accuracy} & \textbf{TTE}
& \textbf{F1-score} & \textbf{Accuracy} & \textbf{TTE}
& \textbf{F1-score} & \textbf{Accuracy} & \textbf{TTE}
& \textbf{F1-score} & \textbf{Accuracy} & \textbf{TTE} \\
\midrule
NND (Flow)
& 0.4305 & 0.4840 & 42.73
& 0.5067 & 0.4847 & 65.36
& 0.4477 & 0.4863 & 59.10
& 0.4817 & 0.4821 & 24.79 \\
NND (Seq.)
& 0.4822 & 0.4180 & 29.69
& 0.5949 & 0.4847 & \textbf{20.39}
& 0.5106 & 0.4366 & 31.31
& 0.5568 & 0.4578 & 24.51 \\
\midrule
PASAD (Sup.)
& 0.7496 & 0.8026 & 84.35
& 0.8567 & 0.8391 & 87.65
& 0.1542 & 0.5281 & 186.75
& 0.8027 & 0.8049 & 118.90 \\
PASAD (Rec.)
& 0.9188 & 0.9259 & 4.85
& 0.9029 & 0.8863 & 49.65
& 0.9295 & 0.9322 & 4.00
& 0.8642 & 0.8584 & 70.90 \\
PASAD (Rec.+CLC)
& 0.9926 & 0.9938 & 6.35
& \textbf{0.9316} & \textbf{0.9248} & 62.90
& 0.9921 & 0.9930 & 6.25
& \textbf{0.9933} & \textbf{0.9930} & 8.40 \\
\midrule
GeCo (Sup.)
& 0.8190 & 0.8282 & 24.35
& 0.8577 & 0.8339 & 64.65
& 0.7769 & 0.7907 & 38.50
& 0.8907 & 0.8723 & 8.40 \\
GeCo (Rec.)
& 0.8593 & 0.8616 & \textbf{2.35}
& 0.8777 & 0.8547 & 35.15
& 0.8796 & 0.8768 & \textbf{2.00}
& 0.8961 & 0.8779 & \textbf{2.65} \\
GeCo (Rec.+CLC)
& \textbf{0.9954} & \textbf{0.9961} & \textbf{2.35}
& 0.8779 & 0.8556 & 37.40
& \textbf{1.0000} & \textbf{1.0000} & \textbf{2.00}
& 0.9725 & 0.9704 & 5.90 \\
\bottomrule
\end{tabular}
\begin{tablenotes}[para]
Note: NND (Flow) and NND (Seq.) use flow-level and packet-sequence features, respectively. Sup. denotes the supervisory-only view; Rec. denotes the view augmented with \tool{}-recovered PVs; Rec.+CLC further applies cross-level consistency checking. S1--S4 are defined in Table~\ref{tab:attack_scenarios}. TTE is measured in seconds, and lower is better. Bold values indicate the best result within each scenario and metric.
\end{tablenotes}
\end{threeparttable}
\end{table*}

\subsubsection{Attack Detection with Recovered Visibility}
\label{sec:security_eval}
Building on the recovered control-level PVs, we evaluate whether the expanded process view improves downstream attack detection. In this evaluation, \tool{} serves as a visibility-recovery front-end for existing attack detection methods, rather than as a standalone detector.

\paragraph{Attack Scenarios and Evaluation Setup.}
Because the available SWaT attack releases do not jointly provide temporally aligned network traffic and historian records required for end-to-end evaluation, we construct aligned attack instances from SWaT normal traces and SWaT attack-prototype values. 
We evaluate attack detection under four representative attack scenarios summarized in Appendix~\ref{app:attack}. These scenarios cover two attack families: control-path deception (S1--S2) and split-view deception (S3--S4), with sensor and actuator targets in each family.

We compare one traffic-based detector and two process-aware detectors. We include \textit{NND}~\cite{schuster2024no} as a traffic-only baseline to represent detection without access to PV values, and report its flow-level and sequence-level configurations. The process-aware detectors are \textit{GeCo}~\cite{wolsing2025gecos} and \textit{PASAD}~\cite{aoudi2018truth}, which operate on PV time series and therefore allow us to evaluate how recovered PVs affect existing process-aware attack detection.

For \textit{GeCo} and \textit{PASAD}, we evaluate two augmentation strategies based on recovered control-level PVs. The baseline setting, \emph{Sup.}, uses only historian-observable supervisory-level PVs. The first augmentation, \emph{Rec.}, adds the recovered control-level PVs to the original supervisory-level PV inputs, so the detector can use both supervisory- and control-level PV evidence. The second augmentation, \emph{Rec.+CLC}, further checks consistency between historian-observable supervisory-level PVs and recovered control-level PVs. High-confidence attack-detection alarms are retained directly, while low-confidence alarms are retained only when corresponding PVs show inconsistent behavior across the supervisory-level and recovered control-level views. Thus, \emph{Rec.} augments detection by adding recovered PV evidence, while \emph{Rec.+CLC} further supports split-view attack detection through cross-level consistency analysis. 
To avoid scenario-specific tuning, all alarm and consistency thresholds are fixed using only the training data and then applied unchanged to all attack instances.

We report F1-score and accuracy to evaluate detection effectiveness. They are computed by comparing attack-detection alarms with the normal/attack labels. We also report time-to-exposure (TTE), defined as the elapsed time from attack onset to the first valid alarm within the attack interval; lower TTE is better.

\begin{figure}[tb]
    \centering
    \includegraphics[width=0.9\linewidth]{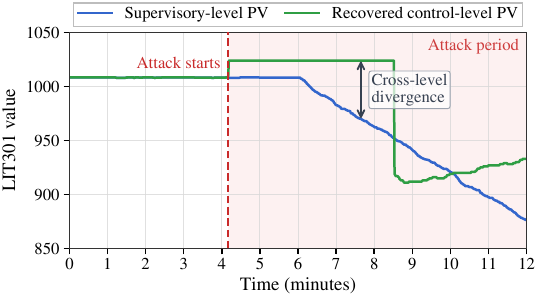}
    \caption{Representative cross-level inconsistency under split-view deception.}
    \label{fig:case}
\end{figure}

\paragraph{Detection Results.}
Table~\ref{tab:security_eval} reports the overall detection results. \textit{NND} provides a traffic-only baseline for detecting physical-process attacks without access to PV values. It sometimes reports early alarms, because flow-level or packet-sequence features can react quickly to traffic perturbations. However, its F1-score and accuracy remain substantially lower than the process-aware detectors across all four scenarios. This suggests that traffic-level evidence alone is insufficient to reliably characterize abnormal process behavior.

For both \textit{PASAD} and \textit{GeCo}, replacing \textit{Sup.} with \textit{Rec.} improves F1-score and accuracy across all four scenarios, while also reducing TTE in most cases. This improvement is obtained by changing the available process evidence rather than the detector itself. Historian records provide a supervisory-level view sampled at a lower frequency, so abnormal process deviations may be delayed or weakened in the recorded PVs. Recovered control-level PVs provide higher-frequency runtime evidence, making these deviations more visible to existing detectors.

Adding cross-level consistency checking further improves detection effectiveness, especially for split-view deception. Under S3 and S4, \textit{PASAD (Rec.+CLC)} improves F1-score from 0.9295 to 0.9921 and from 0.8642 to 0.9933 over \textit{PASAD (Rec.)}, while \textit{GeCo (Rec.+CLC)} improves F1-score from 0.8796 to 1.0000 and from 0.8961 to 0.9725. Figure~\ref{fig:case} illustrates the reason. After the attack starts, the historian record of the attack-target PV remains benign-looking, whereas the recovered control-level PV reflects the attack-side value. This mismatch creates a cross-level divergence that helps confirm low-confidence alarms under split-view deception. \emph{Rec.+CLC} can have higher TTE than \emph{Rec.}, because confirming low-confidence alarms requires waiting until this divergence becomes observable.

\begin{figure}[tb]
    \centering
    \includegraphics[width=0.9\linewidth]{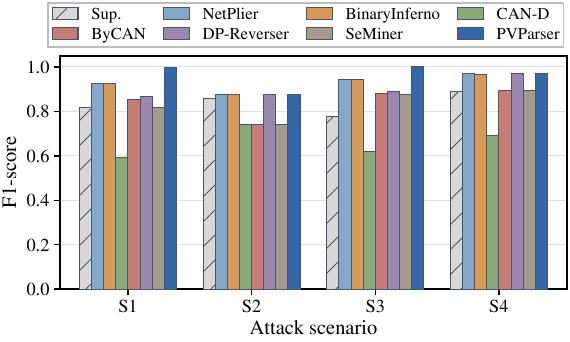}
    \caption{End-to-end F1-scores of GeCo (Rec.+CLC) using control-level PVs recovered by PRE methods across the four attack scenarios. Sup. uses only supervisory-level PVs.}
    \label{fig:pre_detection}
\end{figure}

\paragraph{Impact of PV Recovery Quality.}
To examine how PV recovery quality affects downstream detection, we apply \textit{GeCo (Rec.+CLC)} using the control-level PVs recovered by each PRE method, while retaining the same detector configuration and thresholds. As shown in Figure~\ref{fig:pre_detection}, \tool{} achieves the highest F1-scores in S1 and S3 and matches the best-performing baselines in S2 and S4. DP-Reverser and NetPlier match \tool{} in S2 and S4 because they recover the same attack-relevant control-level PVs. In contrast, DP-Reverser's incorrect \texttt{FIT301} association and NetPlier's omission of this PV respectively reduce their F1-scores in S1 and S3. Baselines with fewer correct recoveries or more incorrect associations provide less useful evidence or introduce misleading inputs to the detector, resulting in lower detection performance. These results demonstrate that downstream security utility depends on both PV recovery coverage and the accuracy of PV--field associations.

Overall, the results show that accurately recovered control-level PVs provide useful additional evidence for existing process-aware detectors and improve detection even when supervisory-level PVs already expose the attacks. Their value is particularly evident when the supervisory- and control-level views diverge.

\section{Related Work}
Prior work related to PV recovery from industrial network traffic can be broadly categorized into active interaction-driven approaches and passive trace-driven approaches~\cite{buscemi2023survey, wu2025reverse}.

Active approaches deliberately perturb the control loop, for example by injecting crafted packets~\cite{luo2024dynpre} or modifying setpoints~\cite{shi2023extracting, ruan2025picacan}, to induce observable communication changes associated with PV updates. In controlled testbeds, ground-truth PV values can then be collected from HMIs~\cite{cai2022seminer} or diagnostic tools~\cite{yu2022towards} and aligned with network traces through timestamp matching and correlation analysis to associate protocol fields with process variables~\cite{verma2021can, lin2024bycan}. However, such approaches are difficult to deploy in operational industrial settings, where arbitrary intervention is often unsafe.

Passive approaches attempt to recover PV structure and semantics from communication traces. Existing methods typically rely on heuristic rules to infer field boundaries~\cite{sun2024variable, qu2025icepre}, syntax~\cite{kleber2022network, qin2024reverse}, and semantics~\cite{zhan2023toward, zhao2024crafting, yang2025patty} from byte- and bit-level patterns. At the protocol-message level, many methods adopt a sequential inference strategy, segmenting messages in a single pass based on heuristic confidence scores~\cite{ye2021netplier, chandler2023binaryinferno, huang2024finebid}. Such strategies make field recovery depend heavily on locally plausible boundary decisions, but local evidence alone may not determine whether a segmentation remains consistent across the whole payload. This limitation becomes especially problematic for PV-carrying payloads, which are often long, deployment-specific, and mixed with non-PV fields, requiring payload-level reasoning rather than isolated boundary decisions.

Overall, existing work falls short of satisfying two key requirements for practical PV recovery from industrial traffic: isolating PV-carrying payloads from heterogeneous industrial traffic and recovering PV fields from long deployment-specific payloads. 
\tool{} addresses these gaps by combining periodic-pattern-based payload localization with search-based field inference that evaluates candidate segmentations at the payload level, enabling automatic PV recovery directly from raw industrial network traffic.

\section{Limitations and Future Work}

\paragraph{Limitations.}
While \tool{} performs well in our evaluation, it has several limitations. First, \tool{} relies on captured network traffic together with historian PV records to recover payload structure and PV semantics. In environments where either source is unavailable or severely incomplete, its applicability may be reduced. Second, \tool{} currently targets unencrypted industrial network traffic, since encryption obscures the byte-level structures and values required for inference. Although some modern deployments adopt encryption, many industrial systems, particularly legacy ones, still transmit traffic between the supervisory level and the control level in plaintext~\cite{lian2024survey}. Third, \tool{} focuses on byte-level industrial protocols operating between the supervisory level and the control level, such as EtherNet/IP, S7comm, and MMS. Bit-level protocols, such as CAN, remain outside the current scope. 
Fourth, \tool{} assumes recurrent PV-carrying payloads with stable byte-level layouts and does not support changed-only or event-driven messages with variable PV lists. In such messages, the same byte positions may correspond to different PVs across payload instances, making field inference and semantic alignment unreliable.

\paragraph{Future Work.}
Future work will proceed in two directions. First, we plan to develop intrinsic PV representations that support field identification and semantic inference even when historian PV records are limited or unavailable. Second, we will extend the search-based inference framework to bit-level industrial protocols, thereby expanding PV recovery to lower-level communication buses and mixed-granularity protocol settings.

\section{Conclusion}
We presented \tool{}, an automated framework for recovering PVs from raw industrial network traffic. \tool{} combines periodic-pattern-based payload localization with search-based field inference to identify PV-carrying payloads and recover their field boundaries and semantics. Experiments on three industrial CPS datasets show that \tool{} achieves over 0.92 accuracy and over 0.96 F1-score for PV field recovery, significantly outperforming six state-of-the-art methods. Further evaluation on SWaT shows that the recovered PVs extend process visibility beyond historian records and improve downstream attack detection. These results demonstrate both the effectiveness of search-based PV recovery and the security value of restoring control-level process visibility.

\section{Acknowledgments}
We thank the anonymous reviewers for their insightful feedback. This work was supported in part by the National Key R\&D Program of China under Grant 2024YFB4709000, the Science and Technology Plan Project of Liaoning Province under Grant 2023JH1/10400076, and the Natural Science Foundation of Liaoning Province under Grant 2026-MS-091.

\bibliographystyle{ACM-Reference-Format}
\bibliography{IEEEabrv, reference}

@STRING{IEEE_J_TII     = "{IEEE} Trans. Ind. Inform."}

@STRING{IEEE_J_SMCS     = "{IEEE} Trans. Syst. Man Cybern. Syst."}

@String{BIT = "{BIT}" }

@String{Computing = "Computing" }

@String{Computer = "{IEEE} Computer" }

@String{Springer = "Springer-Verlag" }

@String{IEEE_J_TII     = "{IEEE} Trans. Ind. Inform."}

@String{IEEE_J_SMCS     = "{IEEE} Trans. Syst. Man Cybern. Syst."}

@article{cai2024ics,
  title={{ICS} anomaly detection based on sensor patterns and actuator rules in spatiotemporal dependency},
  author={Cai, Jun and Wei, Zeheng and Luo, Jianzhen},
  journal={IEEE Transactions on Industrial Informatics},
  volume={20},
  number={8},
  pages={10647--10656},
  year={2024},
  publisher={IEEE}
}

@article{sun2024secure,
  title={Secure Tracking Control and Attack Detection for Power Cyber-Physical Systems Based on Integrated Control Decision},
  author={Sun, Chaowei and Su, Qingyu and Li, Jian},
  journal={IEEE Transactions on Information Forensics and Security},
  volume={20},
  number={},
  pages={968-979},
  year={2024},
  publisher={IEEE}
}

@inproceedings{abbas2024sain,
  title={{SAIN}: Improving {ICS} Attack Detection Sensitivity via {State-Aware} Invariants},
  author={Abbas, Syed Ghazanfar and Ozmen, Muslum Ozgur and Alsaheel, Abdulellah and Khan, Arslan and Celik, Z Berkay and Xu, Dongyan},
  booktitle={33rd USENIX Security Symposium (USENIX Security 24)},
  pages={6597--6613},
  year={2024}
}

@inproceedings{xu2024physcout,
  title={PhyScout: Detecting Sensor Spoofing Attacks via Spatio-temporal Consistency},
  author={Xu, Yuan and Deng, Gelei and Han, Xingshuo and Li, Guanlin and Qiu, Han and Zhang, Tianwei},
  booktitle={Proceedings of the 2024 on ACM SIGSAC Conference on Computer and Communications Security},
  pages={1879--1893},
  year={2024}
}

@inproceedings{fung2024attributions,
  title={Attributions for {ML-based ICS} anomaly detection: From theory to practice},
  author={Fung, Clement and Zeng, Eric and Bauer, Lujo},
  booktitle={Proceedings of the 31st Annual Network and Distributed System Security Symposium, {NDSS}},
  year={2024}
}

@inproceedings{wolsing2025gecos,
  title={GeCos Replacing Experts: Generalizable and Comprehensible Industrial Intrusion Detection},
  author={Wolsing, Konrad and Wagner, Eric and Lux, Luisa and Wehrle, Klaus and Henze, Martin and Fraunhofer, FKIE},
  booktitle={34th USENIX Security Symposium (USENIX Security 25)},
  year={2025}
}

@inproceedings{qin2024reverse,
  title={Reverse engineering industrial protocols driven by control fields},
  author={Qin, Zhen and Yang, Zeyu and Geng, Yangyang and Che, Xin and Wang, Tianyi and Zhu, Hengye and Cheng, Peng and Chen, Jiming},
  booktitle={IEEE INFOCOM 2024-IEEE conference on computer communications},
  pages={2408--2417},
  year={2024},
  organization={IEEE}
}

@article{qu2025icepre,
  title={{ICEPRE}: {ICS} Protocol Reverse Engineering via Data-Driven Concolic Execution},
  author={Qu, Yibo and Fang, Dongliang and Wang, Zhen and Cheng, Jiaxing and Si, Shuaizong and Chen, Yongle and Sun, Limin},
  journal={Proceedings of the ACM on Software Engineering},
  volume={2},
  number={ISSTA},
  pages={2384--2406},
  year={2025},
  publisher={ACM New York, NY, USA}
}

@article{pu2024cormand2,
  title={{CORMAND2}: A deception attack against industrial robots},
  author={Pu, Hongyi and He, Liang and Cheng, Peng and Chen, Jiming and Sun, Youxian},
  journal={Engineering},
  volume={32},
  pages={186--201},
  year={2024},
  publisher={Elsevier}
}

@inproceedings{goh2016dataset,
  title={A dataset to support research in the design of secure water treatment systems},
  author={Goh, Jonathan and Adepu, Sridhar and Junejo, Khurum Nazir and Mathur, Aditya},
  booktitle={International conference on critical information infrastructures security},
  pages={88--99},
  year={2016},
  organization={Springer}
}

@article{lian2024survey,
  title={A Survey on Cyber-Attacks for Cyber-Physical Systems: Modeling, Defense and Design},
  author={Lian, Zhi and Shi, Peng and Chen, Mou},
  journal={IEEE Internet of Things Journal},
  year={2024},
  volume={12},
  number={2},
  pages={1471-1483},
  publisher={IEEE}
}

@article{verma2021can,
  title={{CAN-D}: A modular four-step pipeline for comprehensively decoding controller area network data},
  author={Verma, Miki E and Bridges, Robert A and Sosnowski, Jordan J and Hollifield, Samuel C and Iannacone, Michael D},
  journal={IEEE Transactions on Vehicular Technology},
  volume={70},
  number={10},
  pages={9685--9700},
  year={2021},
  publisher={IEEE}
}

@article{cai2022seminer,
  title={{SeMiner}: Side-information-based semantics miner for proprietary industrial control protocols},
  author={Cai, Jun and Zhong, Weijian and Luo, Jianzhen},
  journal={IEEE Internet of Things Journal},
  volume={9},
  number={22},
  pages={22796--22810},
  year={2022},
  publisher={IEEE}
}

@inproceedings{yu2022towards,
  title={Towards automatically reverse engineering vehicle diagnostic protocols},
  author={Yu, Le and Liu, Yangyang and Jing, Pengfei and Luo, Xiapu and Xue, Lei and Zhao, Kaifa and Zhou, Yajin and Wang, Ting and Gu, Guofei and Nie, Sen and others},
  booktitle={31st USENIX Security Symposium (USENIX Security 22)},
  pages={1939--1956},
  year={2022}
}

@article{lin2024bycan,
  title={Bycan: Reverse engineering controller area network ({CAN}) messages from bit to byte level},
  author={Lin, Xiaojie and Ma, Baihe and Wang, Xu and Yu, Guangsheng and He, Ying and Liu, Ren Ping and Ni, Wei},
  journal={IEEE Internet of Things Journal},
  year={2024},
  volume={11},
  number={21},
  pages={35477-35491},
  publisher={IEEE}
}

@inproceedings{kleber2022network,
  title={Network message field type clustering for reverse engineering of unknown binary protocols},
  author={Kleber, Stephan and Kargl, Frank and State, Milan and Hollick, Matthias},
  booktitle={2022 52nd Annual IEEE/IFIP International Conference on Dependable Systems and Networks Workshops (DSN-W)},
  pages={80--87},
  year={2022},
  organization={IEEE}
}

@inproceedings{ye2021netplier,
  title={{NetPlier}: Probabilistic Network Protocol Reverse Engineering from Message Traces},
  author={Ye, Yapeng and Zhang, Zhuo and Wang, Fei and Zhang, Xiangyu and Xu, Dongyan},
  booktitle={Proceedings of the 28th Annual Network and Distributed System Security Symposium, {NDSS}},
  year={2021}
}

@inproceedings{chandler2023binaryinferno,
  title={{BinaryInferno}: A Semantic-Driven Approach to Field Inference for Binary Message Formats},
  author={Chandler, Jared and Wick, Adam and Fisher, Kathleen},
  booktitle={Proceedings of the 30th Annual Network and Distributed System Security Symposium, {NDSS}},
  year={2023}
}

@article{huang2024finebid,
  title={{FineBID}: Fine-grained Protocol Reverse Engineering for Bit-level Field IDentification},
  author={Huang, Tao and Gao, Yansong and Zheng, Yifeng and Wang, Zhanfeng and Hu, Chao and Fu, Anmin},
  journal={IEEE Transactions on Dependable and Secure Computing},
  year={2024},
  volume={22},
  number={3},
  pages={2670-2686},
  publisher={IEEE}
}

@inproceedings{ahmed2017wadi,
  title={{WADI}: a water distribution testbed for research in the design of secure cyber physical systems},
  author={Ahmed, Chuadhry Mujeeb and Palleti, Venkata Reddy and Mathur, Aditya P},
  booktitle={Proceedings of the 3rd international workshop on cyber-physical systems for smart water networks},
  pages={25--28},
  year={2017}
}

@inproceedings{villa2025icsquartz,
  title={{{ICSQuartz}: Scan Cycle-Aware and Vendor-Agnostic Fuzzing for Industrial Control Systems}},
  author={Villa, Corban and Doumanidis, Constantine and Lamri, Hithem and Rajput, Prashant Hari Narayan and Maniatakos, Michail},
  booktitle={Proceedings of the 32nd Annual Network and Distributed System Security Symposium, {NDSS}},
  pages={1--18},
  year={2025}
}

@inproceedings{lemay2016providing,
  title={Providing {SCADA} network data sets for intrusion detection research},
  author={Lemay, Antoine and Fernandez, Jos{\'e} M},
  booktitle={9th Workshop on Cyber Security Experimentation and Test (CSET 16)},
  year={2016}
}

@inproceedings{deutschmann2024conformal,
  title={Conformal autoregressive generation: Beam search with coverage guarantees},
  author={Deutschmann, Nicolas and Alberts, Marvin and Mart{\'\i}nez, Mar{\'\i}a Rodr{\'\i}guez},
  booktitle={Proceedings of the AAAI Conference on Artificial Intelligence},
  volume={38},
  number={10},
  pages={11775--11783},
  year={2024}
}

@article{buscemi2023survey,
  title={A survey on controller area network reverse engineering},
  author={Buscemi, Alessio and Turcanu, Ion and Castignani, German and Panchenko, Andriy and Engel, Thomas and Shin, Kang G},
  journal={IEEE Communications Surveys \& Tutorials},
  volume={25},
  number={3},
  pages={1445--1481},
  year={2023},
  publisher={IEEE}
}

@inproceedings{katcher2025investigation,
  title={An investigation of interaction and information needs for protocol reverse engineering automation},
  author={Katcher, Samantha and Mattei, James and Chandler, Jared and Votipka, Daniel},
  booktitle={Proceedings of the 2025 CHI Conference on Human Factors in Computing Systems},
  pages={1--21},
  year={2025}
}

@inproceedings{luo2024dynpre,
  title={DynPRE: Protocol reverse engineering via dynamic inference},
  author={Luo, Zhengxiong and Liang, Kai and Zhao, Yanyang and Wu, Feifan and Yu, Junze and Shi, Heyuan and Jiang, Yu},
  booktitle={Proceedings of the 31st Annual Network and Distributed System Security Symposium, {NDSS}},
  pages={1--18},
  year={2024}
}

@inproceedings{shi2023extracting,
  title={Extracting protocol format as state machine via controlled static loop analysis},
  author={Shi, Qingkai and Xu, Xiangzhe and Zhang, Xiangyu},
  booktitle={32nd USENIX Security Symposium (USENIX Security 23)},
  pages={7019--7036},
  year={2023}
}

@article{ruan2025picacan,
  title={{PicaCAN}: Reverse Engineering Physical Semantics of Signals in {CAN} Messages Using Physically-Induced Causalities},
  author={Ruan, Yucheng and Zhao, Chengcheng and Yang, Zeyu and Shu, Yuanchao and Cheng, Peng and Chen, Jiming},
  journal={IEEE Transactions on Mobile Computing},
  year={2025},
  volume={24},
  number={7},
  pages={5871-5887},
  publisher={IEEE}
}

@inproceedings{sun2024variable,
  title={Variable-length Field Extraction for Unknown Binary Network Protocols},
  author={Sun, Xiuwen and Li, Huiying and Chen, Yu and Cui, Jie and Zhong, Hong},
  booktitle={2024 IEEE 49th Conference on Local Computer Networks (LCN)},
  pages={1--7},
  year={2024},
  organization={IEEE}
}

@article{zhan2023toward,
  title={Toward automated field semantics inference for binary protocol reverse engineering},
  author={Zhan, Mengqi and Li, Yang and Li, Bo and Zhang, Jinchao and Li, Chuanrong and Wang, Weiping},
  journal={IEEE Transactions on Information Forensics and Security},
  volume={19},
  pages={764--776},
  year={2023},
  publisher={IEEE}
}

@article{zhao2024crafting,
  title={Crafting Binary Protocol Reversing via Deep Learning With Knowledge-Driven Augmentation},
  author={Zhao, Sen and Yang, Shouguo and Wang, Zhen and Liu, Yongji and Zhu, Hongsong and Sun, Limin},
  journal={IEEE/ACM Transactions on Networking},
  year={2024},
  volume={32},
  number={6},
  pages={5399-5414},
  publisher={IEEE}
}

@article{yang2025patty,
  title={Patty: Pattern Series-Based Semantics Analysis for Agnostic Industrial Control Protocols},
  author={Yang, Daoqing and Yao, Yu and Shan, Yao and Yang, Licheng and Yang, Wei and Liu, Fuyi and Wu, Yunfeng},
  journal={IEEE Transactions on Information Forensics and Security},
  year={2025},
  volume={20},
  number={},
  pages={5478-5491},
  publisher={IEEE}
}

@inproceedings{schuster2024no,
  title={No need for details: effective anomaly detection for process control traffic in absence of protocol and attack knowledge},
  author={Schuster, Franka and K{\"o}nig, Hartmut},
  booktitle={Proceedings of the 27th International Symposium on Research in Attacks, Intrusions and Defenses},
  pages={278--297},
  year={2024}
}

@article{ghazo2024andvi,
  title={{ANDVI}: Automated Network Device and Vulnerability Identification in {SCADA/ICS} by Passive Monitoring},
  author={Ghazo, Alaa T AL and Kumar, Ratnesh},
  journal=IEEE_J_SMCS,
  volume={54},
  number={4},
  pages={2539---2550},
  year={2024},
  publisher={IEEE}
}

@inproceedings{aoudi2018truth,
  title={Truth will out: Departure-based process-level detection of stealthy attacks on control systems},
  author={Aoudi, Wissam and Iturbe, Mikel and Almgren, Magnus},
  booktitle={Proceedings of the 2018 ACM SIGSAC conference on computer and communications security},
  pages={817--831},
  year={2018}
}

@inproceedings{lin2019timing,
  title={Timing patterns and correlations in spontaneous {SCADA} traffic for anomaly detection},
  author={Lin, Chih-Yuan and Nadjm-Tehrani, Simin},
  booktitle={22nd international symposium on research in attacks, intrusions and defenses (RAID 2019)},
  pages={73--88},
  year={2019}
}

@article{sheng2021cyber,
  title={A cyber-physical model for SCADA system and its intrusion detection},
  author={Sheng, Chuan and Yao, Yu and Fu, Qiang and Yang, Wei},
  journal={Computer Networks},
  volume={185},
  pages={107677},
  year={2021},
  publisher={Elsevier}
}

@inproceedings{adepu2018epic,
  title={Epic: An electric power testbed for research and training in cyber physical systems security},
  author={Adepu, Sridhar and Kandasamy, Nandha Kumar and Mathur, Aditya},
  booktitle={International Workshop on Security and Privacy Requirements Engineering},
  pages={37--52},
  year={2018},
  organization={Springer}
}

@inproceedings{shin2020hai,
  title={{HAI} 1.0:{HIL-based} augmented {ICS} security dataset},
  author={Shin, Hyeok-Ki and Lee, Woomyo and Yun, Jeong-Han and Kim, HyoungChun},
  booktitle={13Th USENIX workshop on cyber security experimentation and test (CSET 20)},
  year={2020}
}

@article{taormina2018battle,
  title={Battle of the attack detection algorithms: Disclosing cyber attacks on water distribution networks},
  author={Taormina, Riccardo and Galelli, Stefano and Tippenhauer, Nils Ole and Salomons, Elad and Ostfeld, Avi and Eliades, Demetrios G and Aghashahi, Mohsen and Sundararajan, Raanju and Pourahmadi, Mohsen and Banks, M Katherine and others},
  journal={Journal of Water Resources Planning and Management},
  volume={144},
  number={8},
  pages={04018048},
  year={2018},
  publisher={American Society of Civil Engineers}
}

@article{bailey2012menlo,
  title={The menlo report},
  author={Bailey, Michael and Dittrich, David and Kenneally, Erin and Maughan, Doug},
  journal={IEEE Security \& Privacy},
  volume={10},
  number={2},
  pages={71--75},
  year={2012},
  publisher={IEEE}
}

@article{ur2024process,
  title={Process-aware security monitoring in industrial control systems: A systematic review and future directions},
  author={ur Rehman, Muaan and Bah{\c{s}}i, Hayretdin},
  journal={International Journal of Critical Infrastructure Protection},
  volume={47},
  pages={100719},
  year={2024},
  publisher={Elsevier}
}

@article{kushner2013real,
  title={The real story of stuxnet},
  author={Kushner, David},
  journal={IEEE Spectrum},
  volume={50},
  number={3},
  pages={48--53},
  year={2013},
  publisher={IEEE}
}

@article{case2016analysis,
  title={Analysis of the cyber attack on the Ukrainian power grid},
  author={Case, Defense Use},
  journal={Electricity information sharing and analysis center (E-ISAC)},
  volume={388},
  number={1-29},
  pages={3},
  year={2016},
  publisher={Washington, DC}
}

@article{stouffer2011guide,
  title={Guide to industrial control systems (ICS) security},
  author={Stouffer, Keith and Falco, Joe and Scarfone, Karen and others},
  journal={NIST special publication},
  volume={800},
  number={82},
  pages={16--16},
  year={2011}
}

@inproceedings{yoo2019control,
  title={Control logic injection attacks on industrial control systems},
  author={Yoo, Hyunguk and Ahmed, Irfan},
  booktitle={IFIP International Conference on ICT Systems Security and Privacy Protection},
  pages={33--48},
  year={2019},
  organization={Springer}
}

@inproceedings{urbina2016limiting,
  title={Limiting the impact of stealthy attacks on industrial control systems},
  author={Urbina, David I and Giraldo, Jairo A and Cardenas, Alvaro A and Tippenhauer, Nils Ole and Valente, Junia and Faisal, Mustafa and Ruths, Justin and Candell, Richard and Sandberg, Henrik},
  booktitle={Proceedings of the 2016 ACM SIGSAC conference on computer and communications security},
  pages={1092--1105},
  year={2016}
}

@misc{pvparser_artifact,
  title        = {{PVParser Artifact}},
  year         = {2026},
  howpublished = {\url{https://anonymous.4open.science/r/PVParser-Release/}},
  note         = {Anonymous artifact repository}
}

@article{wu2025reverse,
  title={Reverse engineering of industrial control protocol: A survey},
  author={Wu, Yuheng and Zhang, Zhenyong and Hetu, Zheqiu and Cheng, Xinyu and Cheng, Peng},
  journal={Security and Safety},
  volume={4},
  pages={2025012},
  year={2025},
  publisher={EDP Sciences and CSPM}
}

@article{browne2012survey,
  title={A survey of monte carlo tree search methods},
  author={Browne, Cameron B and Powley, Edward and Whitehouse, Daniel and Lucas, Simon M and Cowling, Peter I and Rohlfshagen, Philipp and Tavener, Stephen and Perez, Diego and Samothrakis, Spyridon and Colton, Simon},
  journal={IEEE Transactions on Computational Intelligence and AI in games},
  volume={4},
  number={1},
  pages={1--43},
  year={2012},
  publisher={IEEE}
}

@article{sheng2026reverse,
  title={Reverse Engineering of Industrial Protocols From Network Traffic},
  author={Sheng, Chuan and Jiang, Shan and Han, Qing-Long and Zhou, Wei and Ma, Wanlun and Zhu, Xiaogang and Wen, Sheng and Xiang, Yang},
  journal=IEEE_J_TII,
  pages={1--13},
  year={2026},
  publisher={IEEE},
  doi={10.1109/TII.2026.3681937}
}

@misc{censys2024,
  author       = {Censys},
  year =         {2024},
  title        = {{The 2024 State of the Internet Report}},
  howpublished = {\url{https://go.censys.com/rs/120-HWT-117/images/2024SOTIR.pdf}},
  note         = {Accessed: Jul. 2, 2026}
}

\section*{Ethical Considerations}

\paragraph{Stakeholders.}
The primary stakeholders in this work are industrial operators, device vendors, security defenders, and the research community. Industrial operators may benefit from improved process visibility, but may also face risk if methods for recovering process variables were misused. Device vendors may benefit from improved understanding of protocol exposure in deployed environments, while also having an interest in protecting proprietary designs and operational confidentiality. Security defenders and researchers may benefit from improved methodology for passive industrial traffic analysis.

\paragraph{Potential Harms and Mitigations.}
Following the Menlo Report~\cite{bailey2012menlo}, we considered two main risks. The first is research-process harm, such as disruption to industrial operation or interference with device behavior during analysis. We mitigate this by restricting \tool{} to passive observation only. It does not inject traffic, modify device behavior, alter controller state, or interfere with industrial processes. 
The second is publication-related misuse, since recovering process-variable semantics from traffic could in principle provide additional process insight for attackers. In practice, however, such misuse requires substantial prerequisites, including access to network traffic, historian PV records, and sufficient contextual knowledge to interpret recovered semantics. We therefore view the risk of opportunistic abuse as limited.

\paragraph{Decision to Proceed and Publish.}
We decided to conduct and publish this research because the expected security benefits outweigh the residual risks. In particular, \tool{} improves visibility into process variables that are often unavailable to existing monitoring systems, thereby supporting anomaly detection, attribution, and process-aware defense in industrial environments. We therefore believe that proceeding with this research and publishing its findings is ethically justified.

\section*{Open Science}

\paragraph{Artifacts Provided.}
To support evaluation of the core methodological and experimental contributions of \tool{}, we provide an anonymous artifact repository containing the implementation of \tool{}, an environment specification, a README with artifact instructions, and recovered outputs produced by the system.

\paragraph{Access During Review.}
The artifact repository is available during double-blind review at: \url{https://anonymous.4open.science/r/PVParser-Release/}. The repository is intended to remain fixed for the duration of the review process.

\paragraph{Artifacts Not Shared and Justification.}
The full industrial datasets used in our experiments are not redistributed in the artifact repository. These datasets originate from publicly released or research-accessible industrial CPS testbeds, and can be obtained from their original providers subject to the corresponding access procedures and usage conditions. We therefore do not re-host or redistribute them in order to respect the distribution policies associated with the original datasets.

\paragraph{Evaluation Scope.}
Although the full datasets cannot be shared, the released implementation, environment specification, README instructions, and recovered outputs enable the program committee to evaluate the core methodology, inferred outputs, and the workflow underlying the reported experiments. In addition, researchers who later obtain legitimate access to the original datasets can further verify the reported results using the released artifacts.

\section*{Generative AI Usage}
The authors used ChatGPT for grammar editing and style polishing. All AI-assisted content was reviewed and revised by the authors, who take full responsibility for the work.

\appendix

\section{Heuristic Rules for Field-Type Inference}
\label{sec:appA}
To guide the MCTS-based inference engine, \tool{} incorporates a lightweight set of domain-informed heuristics that assign soft confidence scores to candidate field types. These heuristics guide the search rather than acting as hard rules, and are designed to capture recurring byte-level and statistical characteristics of industrial process data. Their thresholds and scoring cues are informed by observations from representative industrial datasets, including SWaT~\cite{goh2016dataset}, WADI~\cite{ahmed2017wadi}, BATADAL~\cite{taormina2018battle}, EPIC~\cite{adepu2018epic}, and HAI~\cite{shin2020hai}, together with heuristic insights from BinaryInferno~\cite{chandler2023binaryinferno}.

\subsection{Float Heuristics (Float32 / Float64)}

\tool{} incorporates IEEE~754-based heuristics, drawing on insights from BinaryInferno and adapting them to float values commonly observed in industrial network traffic.

\paragraph{Bit-Structure Ratio (L-Ratio).}
This heuristic evaluates the balance between significand-bit and exponent-bit variability in candidate slices. L-Ratio is computed as
\[
    \text{L-Ratio} =
    \frac{\text{average significand-bit frequency}}
         {\text{maximum exponent-bit frequency}}.
\]
Typical ranges are $0.42$--$0.55$ for Float32 and $0.40$--$0.56$ for Float64. Slices falling far outside these ranges receive lower float confidence.

\paragraph{Constant-Significand Stripes.}
This heuristic evaluates structural entropy in the significand field. If more than $40\%$ of significand-bit positions remain constant across samples, the slice is treated as less likely to encode float values, and its confidence is reduced.

\paragraph{Value-Range Plausibility.}
This heuristic examines whether reconstructed values fall within realistic engineering magnitudes, specifically $10^{-6}$--$10^{6}$ for Float32 and $10^{-8}$--$10^{8}$ for Float64. Extreme values outside these ranges are penalized.

\paragraph{Fractionality.}
This heuristic measures continuity characteristics of candidate values. If at least $70\%$ of values contain fractional components, the slice is treated as more likely to correspond to continuous physical measurements, and float confidence is increased.

\paragraph{Outcome.}
These signals are combined into a float-confidence score in $[0,1]$, where higher scores indicate stronger evidence that the candidate field is float-encoded.

\subsection{Integer Heuristics (8/16/32/64-bit)}

These heuristics characterize integer-encoded PVs in ICS payloads, such as discrete counts, bounded setpoints, and integer-scaled measurements.

\paragraph{Range Validity.}
This heuristic checks whether values remain within the valid signed or unsigned range for the candidate bit width. Out-of-range values reduce integer confidence.

\paragraph{Range Span Utilization.}
This heuristic measures how much of the representable integer range is actually used:
\[
U = \frac{\max(v) - \min(v)}{2^w - 1},
\]
where $w$ is the candidate bit width. Industrial PVs often occupy a narrow operating band despite using wider integer types, and low utilization $U$ is treated as supporting evidence for integer encoding.

\paragraph{Endian Entropy Pattern.}
This heuristic analyzes byte-wise entropy under the assumed endianness. A monotonic decrease in entropy from the least-significant byte to the most-significant byte is treated as evidence for little-endian integer encoding, while a monotonic increase is treated as evidence for big-endian layout.

\paragraph{Temporal Variation Cues.}
This heuristic examines how values evolve over time by measuring first-order differences and dispersion. Consistent and structured changes, such as smooth or steady variation, strengthen evidence for integer-encoded PVs, whereas long periods of unchanged values weaken confidence.

\paragraph{Outcome.}
These signals are combined into an integer-confidence score in $[0,1]$, where higher scores indicate stronger evidence that the candidate field is integer-encoded.

\subsection{String Heuristics (ASCII / UTF-8)}

These heuristics target human-readable labels, tag names, and mode strings commonly seen in ICS protocols.

\paragraph{ASCII Validity.}
This heuristic checks whether values form valid ASCII strings and measures the proportion of printable characters. Higher printable and valid-ASCII ratios increase confidence, while control characters or encoding failures reduce it.

\paragraph{Printable Character Composition.}
This heuristic rewards strings dominated by alphanumeric characters and basic punctuation commonly found in engineering identifiers, and penalizes strings with frequent control or non-printable characters.

\paragraph{Length Reasonableness.}
This heuristic down-weights confidence for very short strings, which are less likely to correspond to meaningful labels or identifiers.

\paragraph{Outcome.}
These signals are combined into a string-confidence score in $[0,1]$, where higher scores indicate stronger evidence that the candidate field is string-encoded.

\section{PV Decoding Output Format}
\label{sec:appB}

Based on the mapping \(\mathcal{M}_{PV}\) in~\eqref{eq:mapping}, \tool{} exports the recovered PV--field correspondences in a machine-readable JSON format. Each entry records a PV identifier together with its inferred byte-level decoding specification, enabling direct extraction of physical process values from network traffic. A representative excerpt is shown in Listing~\ref{lst:pv_json}.

\begin{lstlisting}[frame=single, basicstyle=\ttfamily\small,
                   captionpos=b,
                   caption={Example PV decoding entry},
                   label={lst:pv_json}]
{
  "group_id": "G1",
  "pv_name": "LIT101",
  "sess_key": ["192.168.1.10", "192.168.1.200"],
  "sol_idx": 0,
  "field_spec": {
    "start_pos": 41,
    "length": 4,
    "type": "float32",
    "endian": "little",
    "confidence": 1.0
  }
}
\end{lstlisting}

\section{Parameter Settings}
\label{app:parameters}

Table~\ref{tab:parameters} summarizes the parameter settings of \tool{} used throughout the evaluation. The alignment-weight tuple follows the order of DTW-based trajectory similarity, KL-based distributional similarity, and range consistency.

\begin{table}[t]
\centering
\small
\caption{Parameter settings of \tool{}. Here $\ell$ denotes the slice length in bytes.}
\label{tab:parameters}
\begin{tabular}{p{3.0cm} p{2.9cm} p{1.7cm}}
\toprule
Component & Parameter & Value \\
\midrule
Periodic pattern detection & Autocorrelation threshold $\theta_{acf}$ & 0.7 \\
Payload slicing & Expansion size $\delta$ & 7 bytes \\
Slice inference engine & MCTS budget per slice & $50 \times \ell \times \ln(\ell)$ iterations \\
Slice inference engine & Top-$k$ candidates per endianness & 10 \\
Inference combination & Beam-search width & 20 \\
Inference alignment & Alignment weights & $(0.3, 0.3, 0.4)$ \\
\bottomrule
\end{tabular}
\end{table}



\begin{figure}[t]
    \centering
    \begin{subfigure}[b]{0.42\textwidth}
        \centering
        \includegraphics[width=\linewidth]{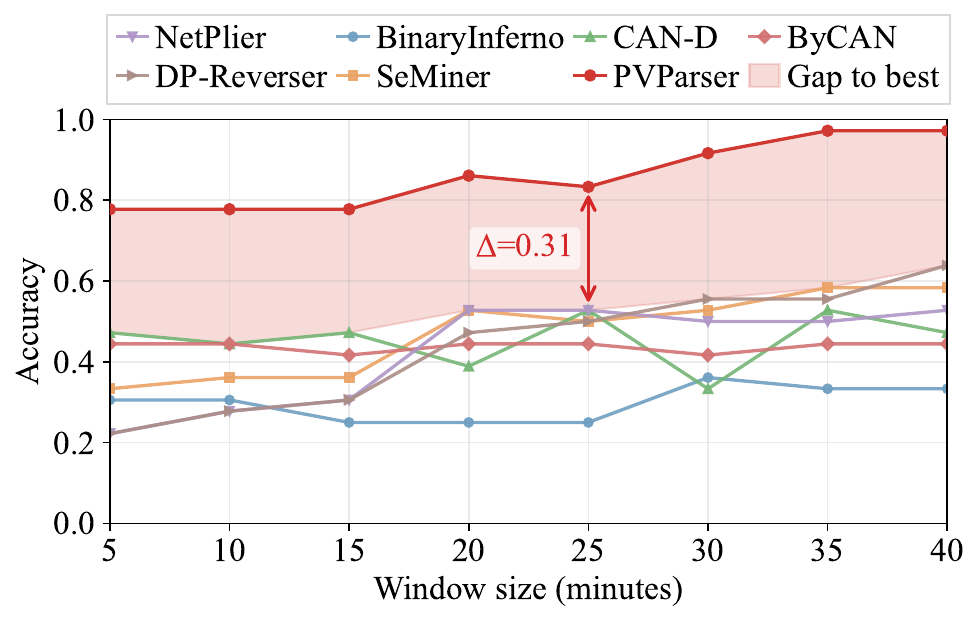}
        \caption{Accuracy}
        \label{fig:operation_state_1}
    \end{subfigure}
    \begin{subfigure}[b]{0.42\textwidth}
        \centering
        \includegraphics[width=\linewidth]{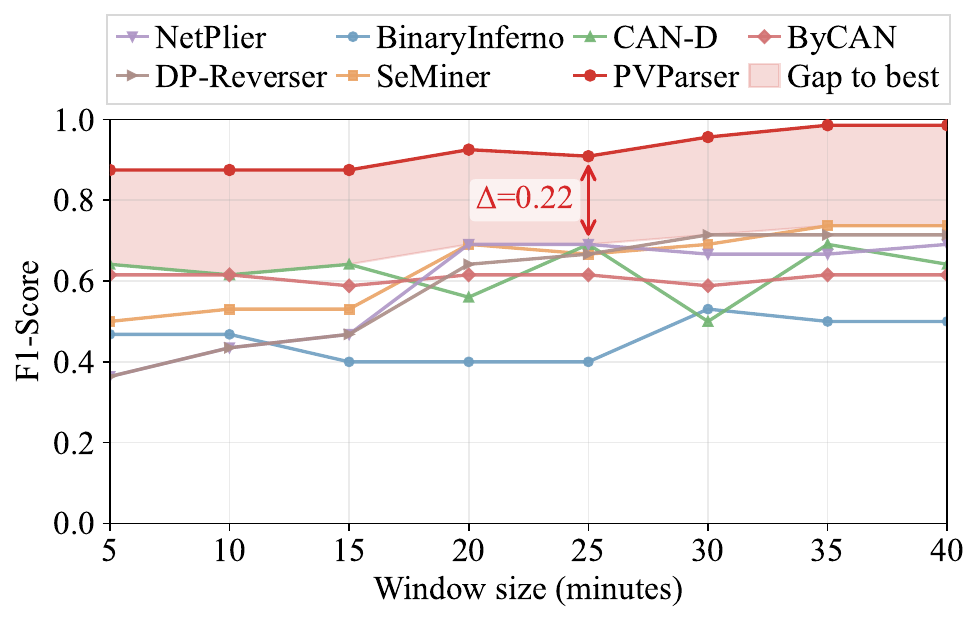}
        \caption{F1-score}
        \label{fig:operation_state_2}
    \end{subfigure}
    \caption{Inference performance across cumulative runtime windows, showing \tool{}'s consistent lead.}
    \label{fig:runtime}
\end{figure}

\section{Impact of Runtime Progression}
\label{sec:window}

To evaluate how runtime progression influences PV field inference, we begin at system startup and gradually expand the observation window in 5-minute increments, cumulatively including all traffic up to each point. For each cumulative window, we perform inference using the same evaluation procedure as in Section~\ref{sec:4.2}. This setup captures inference behavior as the system transitions from startup to steady state.

As shown in Figure~\ref{fig:runtime}, runtime progression clearly affects inference performance. All methods perform worse during the early stage of system operation, and performance stabilizes after approximately 35 minutes. In the earliest windows, many process variables have not yet begun to change, leaving portions of the payload static and difficult to distinguish from non-PV bytes. Short windows also provide fewer samples, weakening value-distribution cues and making inference less reliable. As the cumulative window increases and more PV dynamics become observable, performance gradually improves and then stabilizes. This observation supports the 35-minute training window used in Section~\ref{sec:4.2}.

Despite these challenges, \tool{} maintains a consistent lead across all runtime windows, outperforming all baselines by at least 0.31 in accuracy and 0.22 in F1-score. We also observe that, once stabilized, the baseline methods converge to a similar performance level and show little further improvement with longer runtime. This plateau suggests a shared limitation of traditional PRE methods based on sequential segmentation, where early boundary decisions constrain later inference. In contrast, \tool{} avoids premature commitments and consistently achieves stronger performance across all runtime windows.

\begin{figure}[tb]
    \centering
    \includegraphics[width=0.85\linewidth]{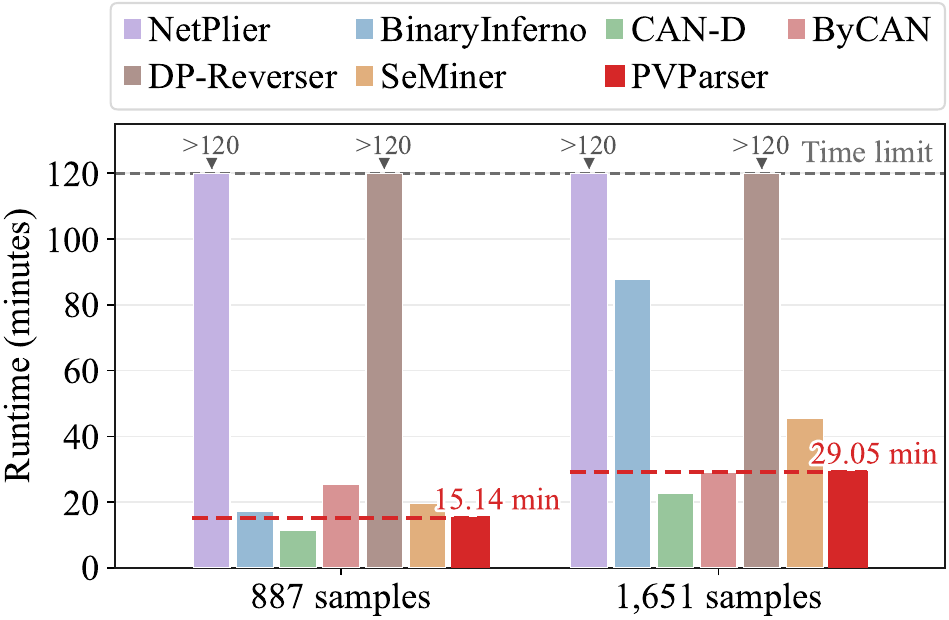}
    \caption{Runtime comparison under two payload sample counts, showing \tool{}'s competitive efficiency.}
    \label{fig:execution_time}
\end{figure}

\begin{table*}[t]
\centering
\caption{Attack scenarios considered in the security evaluation.}
\label{tab:attack_scenarios}
\small
\begin{tabular}{p{1.5cm} p{3.3cm} p{2.4cm} p{2.4cm} p{5.6cm}}
\toprule
\textbf{Scenario} & \textbf{Subtype} & \textbf{Attack target} & \textbf{Injection location} & \textbf{Supervisory-level view} \\
\midrule

\multirow{2}{1.5cm}{Control-path deception}
& \textbf{S1}: Sensor-value deception
& Sensor-value PV
& PLC control path
& \multirow{2}{5.6cm}{Historian records reflect the resulting process consequences.} \\

\cmidrule(lr){2-4}

& \textbf{S2}: Actuator-state deception
& Actuator-state PV
& PLC control path
& \\

\midrule

\multirow{2}{1.5cm}{Split-view deception}
& \textbf{S3}: Sensor-value split-view
& Sensor-value PV
& Dual-path injection
& \multirow{2}{5.6cm}{Historian records contain a benign-looking value for the attack-target PV.} \\

\cmidrule(lr){2-4}

& \textbf{S4}: Actuator-state split-view
& Actuator-state PV
& Dual-path injection
& \\

\bottomrule
\end{tabular}
\end{table*}

\begin{table*}[t]
\centering
\caption{Attack-relevant PV subspaces derived from two representative SWaT attack prototypes.}
\label{tab:attack_subspace}
\small
\begin{tabular}{p{5.4cm} p{2.2cm} p{2.8cm} p{3.4cm} p{1.3cm}}
\toprule
\textbf{Reference attack} & \textbf{Manipulated PVs} & \textbf{Directly affected PVs} & \textbf{Attack subspace} & \textbf{Scenarios} \\
\midrule

LIT301 spoof $\rightarrow$ underflow in T301
& LIT301
& P101, MV201, FIT301
& P101, MV201, FIT301, LIT301
& S1, S3 \\

\midrule

MV201/P101 manipulation $\rightarrow$ overflow in T301
& P101, MV201
& FIT201, LIT301
& P101, MV201, FIT201, LIT301
& S2, S4 \\

\bottomrule
\end{tabular}
\end{table*}

\section{Execution Time Analysis}
\label{sec:efficiency}
We analyze the efficiency of \tool{} from both offline and online perspectives. \tool{} first performs offline recovery to learn the decoding specification, and then uses this learned specification for online payload matching and PV parsing during deployment. The main computational cost lies in the offline recovery stage, while online operation does not repeat the full recovery procedure.

For offline recovery, \tool{} consists of two stages: PV-carrying payload localization and PV field inference. Payload localization runs in \(O(N_s)\), where \(N_s\) is the number of PV-carrying payload samples. This is because it evaluates a bounded number of candidate periods within a fixed window, and each sliding-window scan is linear in \(N_s\). PV field inference runs in \(O(N_s \times L_p)\), where \(L_p\) is the payload length. This is because the total length of all inferred payload slices is approximately \(L_p\), while the inference cost scales linearly with the number of samples. Since runtime is dominated by field inference, the overall complexity of offline recovery is \(O(N_s \times L_p)\).

For empirical evaluation, we compare only the offline PV field inference stage, since all baselines use our PV-carrying payload localization results for fairness. We measure runtime on SWaT historian-PLC\_2 traffic (218-byte payload) under two observation windows: 15 minutes (887 samples) and 30 minutes (1,651 samples). As shown in Figure~\ref{fig:execution_time}, \tool{} completes offline inference in 15.14 minutes and 29.05 minutes under the two settings, exhibiting near-linear growth as the sample volume increases, which is consistent with the theoretical analysis.

\tool{} achieves the second-fastest runtime among all methods. Although CAN-D is marginally faster, its inference capability is substantially weaker, limiting the practical value of this runtime advantage. In contrast, heavyweight approaches such as NetPlier and DP-Reverser exceed the 120-minute time limit, while SeMiner and BinaryInferno scale less favorably as the number of samples grows. Overall, \tool{} offers a strong balance between offline recovery cost and reconstruction quality.

To further examine offline recovery on longer payloads, we evaluate the longest 1,469-byte payload in WADI. \tool{} completes PV field inference in 102.18 minutes under the 15-minute observation window (869 samples) and 197.63 minutes under the 30-minute observation window (1,731 samples). Similar to the 218-byte SWaT payload, the runtime nearly doubles as the sample volume doubles, consistent with the complexity analysis above.

To clarify deployment-time cost, we also measure online parsing after the decoding specification has been learned. In this stage, \tool{} only performs payload matching and PV decoding, without repeating the offline recovery stage. 
Across the two SWaT observation windows, online parsing sustains an average throughput of about 24{,}577 payloads/s, corresponding to about \(40.7\,\mu\text{s}\) per payload. This throughput is well above the observed traffic rate, indicating that \tool{} can support real-time traffic parsing once the offline recovery stage is completed.

In deployment, the learned decoding specification can be reused as long as the communication layout remains unchanged. Its validity can be periodically checked by comparing decoded PV values with the corresponding historian records. Accordingly, the offline recovery stage should be rerun after known configuration, engineering, or firmware changes, or when persistent mismatches indicate that the payload layout has changed.

\section{Details of Attack Detection Evaluation}
\label{app:attack}

This appendix provides additional details on the attack detection evaluation in Section~\ref{sec:security_eval}, including the attack scenarios and attack-instance construction.

\paragraph{Attack Scenarios.}
Table~\ref{tab:attack_scenarios} summarizes the four attack scenarios considered in our evaluation. These scenarios represent physical-process attacks that manipulate sensor values or actuator states carried by industrial communication messages. Such manipulation can be enabled by common ICS attack capabilities, such as man-in-the-middle (MITM) attacks or compromised PLC/gateway components~\cite{yoo2019control, pu2024cormand2}.

To ground the scenarios in realistic process effects, we derive them from two representative SWaT attack prototypes with concrete physical consequences. As summarized in Table~\ref{tab:attack_subspace}, \texttt{LIT301} spoofing can lead to underflow in tank T301, while coordinated manipulation of \texttt{MV201}/\texttt{P101} can lead to overflow in tank T301. For each prototype, we instantiate two cases to evaluate attack detection when historian records either reveal process consequences or do not reveal the abnormal value of the attack-target PV. In control-path deception (S1--S2), the manipulated value affects how the PLC executes its control logic, and historian records reflect the resulting process consequences. In split-view deception (S3--S4), the PLC receives the abnormal value, while historian records contain a benign-looking value for the attack-target PV.

\paragraph{Attack-Instance Construction.}
For the four scenarios, we construct attack instances from SWaT network traffic and historian records collected during normal system operation. The goal is to obtain attack data with aligned network traffic, historian records, and attack labels. We use 30 minutes of normal SWaT data for detector training and the subsequent 30 minutes of normal data as the clean test template. For each scenario, we generate 20 attack instances by sampling attack start times in the clean test template and inserting the corresponding SWaT attack-prototype data into the normal data. The inserted data is adjusted to preserve plausible process-state transitions rather than applying abrupt value replacement. We label the interval before the injection start time as normal and the injection interval as attack. The post-injection recovery period is excluded from evaluation because its ground-truth labels are ambiguous.

For sensor-value scenarios S1 and S3, we manipulate \texttt{LIT301} and replay \texttt{MV201}, \texttt{P101}, and \texttt{FIT301} as affected process responses. For actuator-state scenarios S2 and S4, we manipulate \texttt{MV201} and \texttt{P101}, and replay \texttt{FIT201} and \texttt{LIT301} as affected process responses. For split-view scenarios S3 and S4, we generate paired supervisory-level and control-level records: the supervisory-level records keep the manipulated PVs benign-looking, while the control-level records retain their attack-side values.


\end{document}